\documentclass[twocolumn]{aastex631}
\usepackage{subfigure,soul}
\usepackage[normalem]{ulem}
\usepackage{graphicx}
\usepackage{bm}
\usepackage{amsthm,amsmath,amssymb}
\usepackage{CJK}
\begin{document}
\begin{CJK*}{UTF8}{gbsn}
\title{The dependence of the structure of planet-opened gaps in protoplanetary disks on radiative cooling}

\author[0009-0006-5706-0364]{Minghao Zhang (张铭浩)}
\affiliation{National Astronomical Observatories, Chinese Academy of Sciences, Beijing 100012, People's Republic of China; \url{zhangminghao23@mails.ucas.ac.cn}}
\affiliation{School of Astronomy and Space Science, University of Chinese Academy of Sciences, Beijing 100049, People's Republic of China}
\author[0000-0002-7575-3176]{Pinghui Huang (黄平辉)}
\affiliation{Department of Physics and Astronomy, University of Victoria, Victoria, BC V8P 5C2, Canada; \url{pinghuih@uvic.ca}, \url{rbdong@uvic.ca}}
\author[0000-0001-9290-7846]{Ruobing Dong (董若冰)}
\affiliation{Department of Physics and Astronomy, University of Victoria, Victoria, BC V8P 5C2, Canada; \url{pinghuih@uvic.ca}, \url{rbdong@uvic.ca}}

\begin{abstract}
Planets can excite density waves and open annular gas gaps in protoplanetary disks. The depth of gaps is influenced by the evolving angular momentum carried by density waves. While the impact of radiative cooling on the evolution of density waves has been studied, a quantitative correlation to connect gap depth with the cooling timescale is lacking.
To address this gap in knowledge, we employ the grid-based code Athena++ to simulate disk-planet interactions, treating cooling as a thermal relaxation process. We establish quantitative dependences 
of steady-state gap depth (Eq. \ref{K3}) and width (Eq. \ref{width_formula}) on planetary mass, Shakura-Sunyaev viscosity, disk scale height, and thermal relaxation timescale $(\beta)$.
We confirm previous results that gap opening is the weakest when thermal relaxation timescale is comparable to local dynamical timescale. Significant variations in gap depth, up to an order of magnitude, are found with different $\beta$. 
In terms of width, a gap is at its narrowest around $\beta=1$, approximately $10\%$ to $20\%$ narrower compared to the isothermal case. When $\beta\sim100$, it can be $\sim20\%$ wider, and higher viscosity enhances this effect. 
We derive possible masses of the gas gap-opening planets in AS 209, HD 163296, MWC 480, and HL Tau, accounting for the uncertainties in local thermal relaxation timescale. 
\end{abstract}


\section{Introduction} \label{sec:intro}
The gravitational perturbation induced by an orbiting planet within a protoplanetary disk (PPD) excites density waves and triggers a redistribution of material in the disk \citep{2022arXiv220309595P}. In the presence of a sufficiently massive planet, the surrounding material in its orbit can be effectively cleared, resulting in the formation of a gap.

The accurate characterization of gaps in PPDs holds significant importance due to its impact on various crucial physical quantities. For instance, the migration rate of a planet within a disk is governed by the planetary gravitational torque, sensitive to the surface density in the gap ($\Sigma_{\rm gap}$) \citep{2018ApJ...861..140K,2022arXiv220309595P}. Likewise, the accretion rate onto a planet is also influenced by $\Sigma_{\rm gap}$ \citep{2021MNRAS.500.2822C}.
Furthermore, a quantitative correlation between the planet mass, and the gap depth and width, can be employed to infer the former when the latter is measured \citep{2017PASJ...69...97K,Zhang_2018}

Previous studies have made significant progresses in exploring the structure of gaps opened by planets through 2D and 3D numerical simulations as well as semi-analytical models. These investigations have explored the evolution of angular momentum and the dependence of gap depth and width on disk and planetary characteristics, including the planet-star mass ratio ($q\equiv M_p/M_\star$), viscosity ($\alpha$), and disk aspect ratio ($h_p$). 
\cite{2014ApJ...782...88F} derived the analytical expression for gap depth by balancing viscous torque and one-sided planetary torque, providing $\Sigma_{\rm gap}/\Sigma_0\sim q^2\alpha^{-1} h_p^{-5}$. This relation was validated by numerical simulations and initiated parameter explorations in the gap-opening process.
Subsequent studies have expanded upon these findings. \cite{2015MNRAS.448..994K,2015ApJ...806L..15K} and \cite{2015ApJ...807L..11D} developed 1D semi-analytical models for steady-state gap structures, presenting more accurate formulas for gap depth based on angular momentum conservation. When considering massive planets, nonlinear effects become important in calculating excitation torques. 
\cite{2016PASJ...68...43K} proposed that the mass of a planet in a PPD can be constrained based on gap width.
\cite{2017PASJ...69...97K} investigated the gap-opening process of deep gaps and angular momentum deposition using numerical simulations and semi-analytical models, incorporating nonlinear density waves.
Furthermore, \cite{2017PASJ...69...97K} defined the cumulative deposited torque ($\Gamma_d$) to study the ability of density waves to exchange angular momentum with local substances and proposed an iterative method for generating gap profiles in the nonlinear regime, reducing the deviation between numerical gap depth and previous analytic formulas.

In addition to these advancements, other explorations have yielded significant results. For example, \cite{Duffell_2015} compared gap depth and width between single and multiple planets, finding that multiple planets of the same mass created wider but shallower gaps. \cite{Fung_2016} conducted 3D numerical simulations, demonstrating that 3D and 2D simulations produced similar gap structures. 
Combining hydrodynamics simulations with radiative transfer, \cite{2017ApJ...835..146D} shows that the observed gap depth and width can constrain the quantity $M_p^2/\alpha$.
\cite{Duffell_2020} derived an empirical formula for rapidly generating synthetic disk profiles, which shows excellent agreement with simulations across a wide range of parameters.
Recent research by \cite{2023MNRAS.518..439S} examined the relationship between gap depth and planetary orbit eccentricity.

However, most previous works on gaps opened by planets in PPDs adopted an isothermal equation of state (EOS). In recent years, the importance of thermodynamics has been emphasized. 
The pioneering work of \cite{2020ApJ...892...65M} first conducted linear theory and numerical simulations to reveal the importance of thermodynamics in understanding the evolution of density waves in accretion disk systems.
Subsequently, \cite{2020MNRAS.493.2287Z} conducted numerical investigations to elucidate the behavior of density waves under different thermal relaxation timescales.  
\cite{2020ApJ...904..121M} investigated the impact of in-plane radiative diffusion by introducing a combined cooling timescale from the effects of both surface and in-plane cooling.
Furthermore, \cite{2022MNRAS.514.1733H} explored the effect of cooling on the Rossby wave instability (RWI) in PPDs through linear analysis. \cite{2023ApJ...943..175W} initiated studies on how thermodynamics affects the accretion of binary systems using numerical simulations.
In addition to the parameterized $\beta-$cooling method mentioned above, there are also studies that adopt realistic terms for heating and cooling in the energy equation \citep{2020A&A...633A..29Z,2023arXiv230514415Z}. Non-negligible discrepancies have been identified between the former approach and the latter \citep{2023arXiv230514415Z}.

In this paper, we study the influence of radiative cooling on the depth, width, and internal structures of gaps opened by giant planets in PPDs. 
When searching for planets in PPDs, our models can provide more precise information about the properties of embedded planets. The lack of such a model (or even an empirical formula) hampers the determination of planet-induced gaps. 
We introduce a thermal-relaxation treatment with a constant timescale to investigate its impact on the gap-opening process. While one planet may open multiple gaps \citep{Dong_2017,Dong_2018,2017ApJ...850..201B,Bae_2018}, we focus on the primary gap around the planet's orbit, mainly because the viscosity utilized in our investigation generally exceeds the typical viscosity associated with multiple gap opening. Gaps form when density waves transfer angular momentum to the local disk, and the depth of the gap depends on the evolution of angular momentum carried by these waves. Considering a non-isothermal EOS leads to distinct dissipation processes for density waves and different gap structures compared to the locally isothermal condition.

The paper is structured as follows.
In Section \ref{sec:numerical}, we present the setups and outcomes of a numerical parameter exploration focusing on the gap-opening process.
Section \ref{sec:gapdepth} discusses the propagation of density waves and the evolution of torques and angular momentum.
Furthermore, we derive empirical formulas for gap depth and width as a function of planetary mass $q$, viscosity coefficient $\alpha$, disk aspect ratio $h_p$, and cooling time scale $\beta$ in the quasi-steady state.
Section \ref{sec:discussion} contains discussions and applications to observed gas gaps related to the obtained results.

\section{Numerical simulations}
\label{sec:numerical}
\subsection{Method} 
To investigate the structure of steady-state gaps opened by planets in disks with radiative cooling, we employ the grid-based code Athena++ \citep{Stone_2020} to solve the conservative equations for mass, momentum, and energy. These equations governing the behavior of the system are expressed in a polar coordinate system $(R, \phi)$:

The continuity equation:
\begin{equation}
\frac{\partial\Sigma}{\partial t}+\nabla\cdot(\Sigma\textbf{\emph{v}})=0.
\end{equation}

The momentum equation:
\begin{equation}
\frac{\partial(\Sigma\textbf{\emph{v}})}{\partial t}+\nabla\cdot(\Sigma\textbf{\emph{v}}\textbf{\emph{v}}+P\mathcal{I})=-\Sigma\nabla\Phi-\nabla\cdot\mathcal{T}_{\text{vis}}.
\end{equation}

The energy equation:
\begin{equation}
\frac{\partial E}{\partial t}+\nabla\cdot[(E+P)\textbf{\emph{v}}]=-\Sigma\textbf{\emph{v}}\cdot\nabla\Phi-\nabla\cdot(\mathcal{T}_{vis}\textbf{\emph{v}})+
\left(\frac{\partial \epsilon}{\partial t}\right)_{\rm cool}
\end{equation}

In these equations, $\Sigma$ represents the surface density of the disk, $\textbf{\emph{v}}$ is the velocity, $P$ denotes the isotropic pressure, and $E$ is the total energy. The term $\mathcal{I}$ represents the identity tensor.
The viscous stress tensor is given by:
\begin{equation}
\mathcal{T}_{vis,ij}=-\Sigma\nu\left(\frac{\partial v_i}{\partial x_j}+\frac{\partial v_j}{\partial x_i}-\frac{2}{3}\frac{\partial v_k}{\partial x_k}\delta_{ij}\right)
\end{equation}

Here, $\nu = \alpha c_sH$ is the kinematic viscosity, where $\alpha$ is the Shakura-Sunyaev viscosity parameter and $c_s$ is the sound speed. The gravitational potential is given by 
$\Phi(R,\phi,t)=
\Phi_\star(R)
+\Phi_p(R,\phi,t)$, where 
$
\Phi_\star(R) 
$
represents the stellar gravitational potential and $\Phi_p(R,\phi,t)$ represents the planetary gravitational potential. 
We adopt the adiabatic equation of state, where the total energy $E$ is defined as 
$E=\frac{1}{2}\Sigma v^2+P/(\gamma-1)$, 
with the adiabatic index $\gamma=1.4$.
Lastly, the term $\left(\frac{\partial \epsilon}{\partial t}\right)_{\text{cool}}$ represents the thermal relaxation term, which will be discussed in more detail later in the paper.

\subsection{Setups}
\subsubsection{Initial and boundary conditions}
Our 2D simulation domain ranges from $R_{\rm min}=0.2$ to $R_{\rm max}=2.4$ radially (in units where the planet-star distance $R_p=1$), and from $0$ to $2\pi$ in azimuth. 
The aspect ratio of the disk is 
\begin{equation}
h(R)=H/R=h_p(R/R_p)^{\frac{1}{4}},
\label{h_over_r}
\end{equation}
where $H$ is the disk scale height. 
The initial surface density and temperature profiles of the disk are
\begin{equation}
\Sigma_{\rm init}=\Sigma_0\left(\frac{R}{R_p}\right)^{-1},
\end{equation}
\begin{equation}
T_{\rm init}=T_0\left(\frac{R}{R_p}\right)^{-1/2},
\end{equation}
where $\Sigma_0$ and $T_0$ are the unperturbed values at $R_p$, set to $1$ and $h_p^2$. 
The initial velocity profiles are the sub-Keplerian velocity modified by a radial pressure gradient along the azimuthal direction:
\begin{equation}
v_{\phi,\rm init}=\sqrt{1-\frac{3}{2}h_p^2(R/R_p)^{-1}}\sqrt{\frac{GM_\star}{R}},
\end{equation}
and the velocity of viscous inflow along the radial direction:
\begin{equation}
v_{R,\rm init}=-\frac{3}{2}\alpha \left(\frac{H}{R}\right)^2\sqrt{\frac{GM_\star}{R}}.
\end{equation}

The boundary conditions are set to match the initial conditions at the respective inner and outer ghost cells, which are positioned outside the computational mesh. Additionally, we incorporate a wave damping zone near the inner and outer boundaries to mitigate unphysical wave reflections at these boundaries~\citep{DeValBorro2006}.

\subsubsection{Gravitational terms}
To mitigate the initial strong shock from the massive planet, we gradually grow the planetary mass to its full extent $M_p$ \citep{2017ApJ...850..201B}: 

\begin{equation}\begin{aligned}
M_{p,t}=M_p\sin\left(\frac{\pi}{2}\frac{t}{t_{\rm grow}}\right) \quad (t<t_{\rm grow})
\end{aligned}\end{equation}

Here, $t_{\rm grow}$ represents the growth time:
\begin{equation}\begin{aligned}
t_{\rm grow}=\min\left(20\frac{M_p}{M_{th}}t_p,100\frac{M_p}{M_{th}}\right),
\end{aligned}\end{equation}
where $t_p = 2\pi\Omega_p^{-1}$ is the planetary orbital period, with $\Omega_p=\Omega(R_p)$.
The gravitational potential of the planet is denoted by $\Phi_p(R,\phi,t)$, 
\begin{equation}
\Phi_p(R,\phi,t)=-\frac{GM_p}{\left[R^2+R_p^2-2RR_p\cos(\phi-\Omega_pt)+R_s^2\right]^{1/2}},
\end{equation}
where $R_p$ and $\Omega_p$ represent the radial position and orbital frequency of the planet, respectively, and $R_s=0.6R_H$ is the softening length.
Here, $R_H=R_p\left(\frac{M_p}{3M_*}\right)^{1/3}$ is the Hill radius of the planet. 
The gravitational potential of the star is denoted as $\Phi_\star(R,\phi,t)=-GM_\star/R$.
If the planet has a significant mass, the center of mass is shifted away from the star. To maintain the star as the origin of the co-rotating non-inertial reference frame, we introduce the indirect potential 
\begin{equation}\begin{aligned}
\Phi_{\rm Ind}&=\frac{GM_pR\cos(\phi-\phi_p)}{R_p^2}
\end{aligned}\end{equation} 
as an extra correction of the effective gravitational potential \citep{2014ApJ...782...88F,Fung_2016}. 

\subsubsection{Thermal relaxation term}
A common approach to treat cooling and heating is to introduce thermal relaxation with a characteristic timescale $t_{\rm cool}$ to achieve a background temperature profile the same as the initial condition \citep{Zhu_2015,2020MNRAS.493.2287Z,2022arXiv220309595P}. We follow this practice and introduce thermal relaxation as:
\begin{equation}
\left(\frac{\partial \epsilon}{\partial t}\right)_{\rm cool}
=-\frac{\Sigma}{\gamma-1}\frac{T-T_{\rm init}}{t_{\rm cool}}
\label{cooling}
\end{equation}
In the above equation, the cooling timescale $t_{\rm cool}$ can be conveniently expressed in a non-dimensional form as 
$t_{\rm cool}=\beta\Omega_K^{-1}$
, where $\Omega_K$ represents the Keplerian angular velocity and $\beta$ is the dimensionless cooling timescale. We treat $\beta$ as a constant in each simulation that does not vary with time or space.

\subsection{Simulation Runs}
Athena++ provides two reconstruction methods, piecewise linear method (PLM, $xorder = 2$), and piecewise parabolic method (PPM, $xorder = 3$). We found no significant difference in the steady state gap profiles using these two methods (see Figure \ref{resolution_test}). As PLM is much less computation time cost, we adopt it as the default reconstruction method in this work.
In all simulations, we selected the Riemann solver HLLC and the Van Leer 2 (VL2) as the time integrator.

\subsubsection{Parameter Space}
The main simulation parameters are summarized in Table.\ref{table:setups}.  Specifically, we vary\\
\begin{itemize}
\item[$\bullet$]the planet-star mass ratio $q$ from $3\times10^{-4}$ to $3\times10^{-3}$,
\item[$\bullet$]the viscosity parameter $\alpha$ from $3\times10^{-4}$ to $1\times10^{-2}$,
\item[$\bullet$]the disk aspect ratio at the planet location $h_p$ from $0.05$ to $0.1$, and
\item[$\bullet$]the cooling time scale $\beta$ from $0.01$ to $100$.
\end{itemize}

\begin{deluxetable*}{cccccclc}
\tablewidth{\textwidth}
\tabcolsep=0.15cm
\tablecaption{Numerical setups\label{table:setups}}
\tabletypesize{\scriptsize}
\tablehead{
\colhead{Run} & \colhead{Planet-star} & \colhead{Viscosity} & \colhead{Disk aspect} & \colhead{Cooling  } & \colhead{Gap-opening} & \colhead{Grid} & \colhead{Running}\\ 
\colhead{   } & \colhead{mass ratio} & \colhead{          } & \colhead{ratio      } & \colhead{timescale} & \colhead{parameter} & \colhead{resolution} & \colhead{time}\\ 
\colhead{} & \colhead{$q$} & \colhead{$\alpha$} & \colhead{$h_p$} & \colhead{$\beta$} & \colhead{$K=q^2\alpha^{-1}h_p^{-5}$} & \colhead{$N_R\times N_\phi$} & \colhead{$t_{\rm run}/t_p$}
}
\startdata
\multicolumn{8}{c}{Regular Runs}\\
\hline
 h5-1 & $3\times10^{-4}$  & $3\times10^{-4}$  &    0.05    &$(0.01,0.1,1,10,100)$ &  960        & $384\times1152$                   &$5000$ \\
 h5-2 & $3\times10^{-4}$  & $1\times10^{-3}$  &    0.05    &$(0.01,0.03,0.1,0.3,1,3,10,30,100)$ &  288       &     $384\times1152$  &$4500$ \\
 h5-3 & $1\times10^{-3}$  & $3\times10^{-4}$  &    0.05    &$(0.01,0.03,0.1,0.3,1,3,10,30,100)$ &  10667     &   $384\times1152$    &$5000$ \\
h5-4 & $1\times10^{-3}$  & $1\times10^{-3}$  &    0.05    &$(0.01,0.1,1,10,100)$ &  3200       &   $384\times1152$                  &$6000$ \\
 h5-5 & $1\times10^{-3}$  & $3\times10^{-3}$  &    0.05    &$(0.01,0.1,1,10,100)$ &  1067       &   $384\times1152$                 &$5000$ \\
 h5-6 & $3\times10^{-3}$  & $1\times10^{-3}$  &    0.05    &$(0.01,0.1,1,10)$ &  28800      &   $384\times1152$                 &$5000$ \\
 h5-7 & $3\times10^{-3}$  & $3\times10^{-3}$  &    0.05    &$(0.01,0.1,1,10,100)$  &  9600      &   $384\times1152$                 &$4000$ \\
 \hline
 h8-1  & $3\times10^{-4}$  & $1\times10^{-4}$  &    0.08    &$(0.01,0.1,1,10,100)$  &  275     &   $384\times1152$                  &$20000$ \\
 h8-2 & $3\times10^{-4}$  & $3\times10^{-4}$  &    0.08    &$(0.01,0.03,0.1,0.3,1,3,10,30,100)$  &  92    &   $384\times1152$       &$4000$ \\
 h8-3 & $1\times10^{-3}$  & $1\times10^{-4}$  &    0.08    &$(0.01,0.1,1,10,100)$  & 3052       &   $384\times1152$                &$20000$ \\
 h8-4 & $1\times10^{-3}$  & $3\times10^{-4}$  &    0.08    &$(0.01,0.03,0.1,0.3,1,3,10,30,100)$  & 1017      &   $384\times1152$     &$5000$ \\
 h8-5 & $3\times10^{-3}$  & $3\times10^{-4}$  &    0.08    &$(0.01,0.1,1,10,100)$  &  9155      &   $384\times1152$                 &$5000$ \\
 h8-6 & $3\times10^{-3}$  & $1\times10^{-3}$  &    0.08    &$(0.01,0.03,0.1,0.3,1,3,10,30,100)$  &  2747     &   $384\times1152$    &$5000$ \\
 h8-7 & $3\times10^{-3}$  & $3\times10^{-3}$  &    0.08    &$(0.01,0.1,1,10,100)$  &  916     &   $384\times1152$                   &$4000$ \\
 \hline
 h10-1 & $3\times10^{-4}$  & $1\times10^{-4}$  &    0.1    &$(0.01,0.1,1,10,100)$  &   90         &   $384\times1152$               &$20000$ \\
 h10-2 & $3\times10^{-4}$  & $3\times10^{-4}$  &    0.1    &$(0.01,0.03,0.1,0.3,1,3,10,30,100)$  &   30        &   $384\times1152$  &$5000$ \\
 h10-3 & $3\times10^{-4}$  & $1\times10^{-3}$  &    0.1    &$(0.01,0.03,0.1,0.3,1,3,10,30,100)$  &    9         &   $384\times1152$ &$5000$ \\
 h10-4 & $1\times10^{-3}$  & $3\times10^{-4}$  &    0.1    &$(0.01,0.1,1,10,100)$  &  333         &  $384\times1152$                &$6000$ \\
 h10-5 & $1\times10^{-3}$  & $1\times10^{-3}$  &    0.1    &$(0.01,0.03,0.1,0.3,1,3,10,30,100)$  &  100       &   $384\times1152$   &$5000$ \\
 h10-6 & $1\times10^{-3}$  & $3\times10^{-3}$  &    0.1    &$(0.01,0.1,1,10,100)$  &  33          &   $384\times1152$               &$4500$ \\
 h10-7 & $3\times10^{-3}$  & $1\times10^{-3}$  &    0.1    &$(0.01,0.1,1,10,100)$  &  900          & $384\times1152$                &$5000$ \\
 h10-8 & $3\times10^{-3}$  & $3\times10^{-3}$  &    0.1    &$(0.01,0.03,0.1,0.3,1,3,10,30,100)$  &  300         & $384\times1152$   &$4500$ \\ 
 \hline
 \multicolumn{8}{c}{Resolution Tests}\\
\hline
 RT128B01-2 & $1\times10^{-3}$  & $1\times10^{-3}$  &    0.05    &$0.1$  &   \nodata      &  $128\times384$                               &$4000$\\ 
 RT256B01-2 & $1\times10^{-3}$  & $1\times10^{-3}$  &    0.05    &$0.1$  &   \nodata      &  $256\times768$                               &$4000$\\ 
 RT384B01-2 & $1\times10^{-3}$  & $1\times10^{-3}$  &    0.05    &$0.1$  &   \nodata      &   $384\times1152$                             &$4000$\\ 
 RT512B01-2 & $1\times10^{-3}$  & $1\times10^{-3}$  &    0.05    &$0.1$  &   \nodata      &   $512\times1536$                             &$4000$\\ 
 \hline
  RT128B100-2 & $1\times10^{-3}$  & $1\times10^{-3}$  &    0.05   &$100$  &   \nodata     &  $128\times384$                               &$4000$\\ 
 RT256B100-2 & $1\times10^{-3}$  & $1\times10^{-3}$  &    0.05    &$100$  &   \nodata     &  $256\times768$                               &$4000$\\ 
 RT384B100-2 & $1\times10^{-3}$  & $1\times10^{-3}$  &    0.05    &$100$  &   \nodata     &   $384\times1152$                             &$4000$\\ 
 RT512B100-2 & $1\times10^{-3}$  & $1\times10^{-3}$  &    0.05    &$100$  &   \nodata     &  $512\times1536$                              &$4000$\\ 
 \hline                 
 RT384B01-3 & $1\times10^{-3}$  & $1\times10^{-3}$  &    0.05    &$0.1$  & \nodata        &   $384\times1152$                             &$4000$\\ 
 RT512B01-3 & $1\times10^{-3}$  & $1\times10^{-3}$  &    0.05    &$0.1$  &  \nodata       &   $512\times1536$                             &$4000$
\enddata
\end{deluxetable*}

\subsubsection{Convergence Tests}
\begin{figure*}[ht]
\centering
\includegraphics[width=\textwidth]{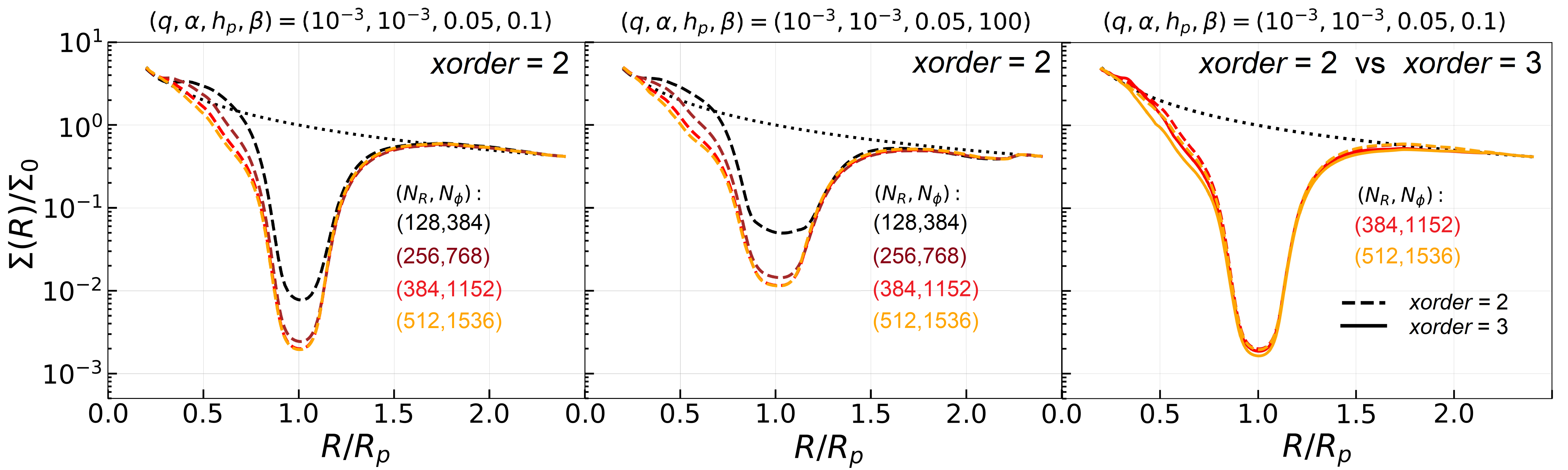}
\caption{\label{resolution_test} 
Radial surface density profiles in simulations with different resolutions and spatial difference orders $xorder$.
The disk and planet parameters are listed on the top.
The left two panels have $xorder=2$ and right panel shows comparison between $xorder=2$ and $xorder=3$. 
}
\end{figure*}
The resolution convergence tests are shown in Figure \ref{resolution_test}. 
We find that simulations with both $\beta=1$ and $\beta=100$ converge at the grid resolution $(N_R, N_\phi)=(384,1152)$, and adopt this as the default. 
With this resolution, approximately 9 grid cells per scale height are resolved at the planet location for $h_p=0.05$, and around 17 grid cells per scale height for $h_p=0.1$.

\begin{figure}[ht]
\centering
\includegraphics[width=0.48\textwidth]{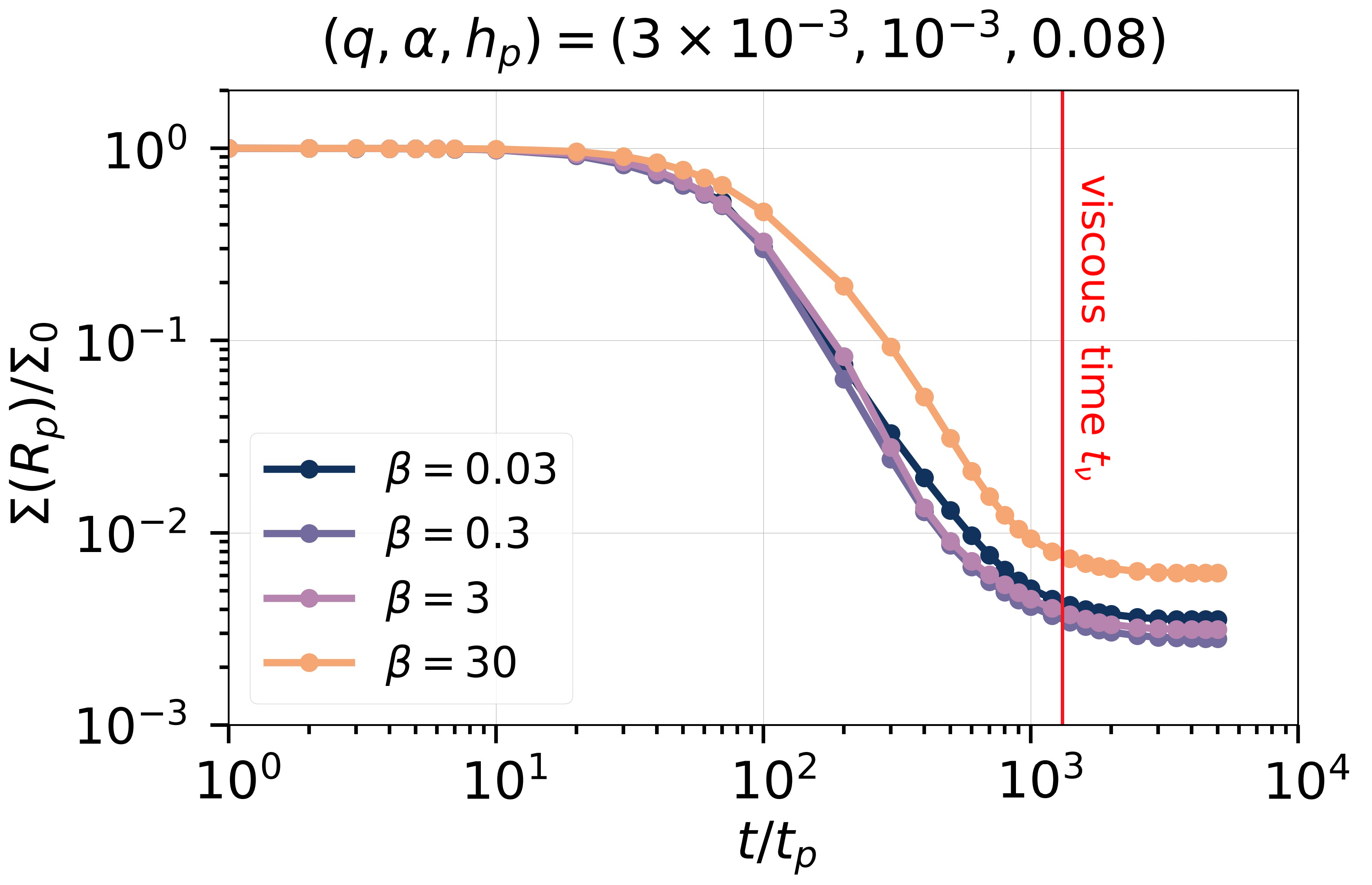}
\caption{\label{time_convergence} 
Azimuthally averaged surface density at the planetary orbit, $\Sigma(R_p)$, as a function of time for 4 simulations with $(q,\alpha,h_p)=(3\times10^{-3},10^{-3},0.08)$ and different $\beta$. 
The viscous time, as given by Eq. (\ref{t_nu}), is marked with a red vertical line.
}
\end{figure}

Each simulation runs until the surface density inside the gap reaches a quasi-steady state. 
We employ the relation, 
\begin{equation}
    t_{\nu}=\frac{(w_{\rm gap}/(2R_p))^2R_p^2}{\nu},
    \label{t_nu}
\end{equation}
as proposed in \cite{2018ApJ...861..140K} to estimate the viscous gap-opening timescale $t_{\nu}$, assuming the gap width $w_{\rm gap}/R_p\approx5.8h_p$ \citep{2017ApJ...835..146D}.
Considering substance in the gap $(R\approx R_p)$, $t_\nu$ is approximately 
$1.3t_p/\alpha$.
For each model, the total running time $t_{\rm run}$, greater than $t_\nu$ in all cases, is listed in Table \ref{table:setups}.
Figure \ref{time_convergence} shows how $\Sigma(R_p)$ varies with time in a run with $(q,\alpha,h_p)=(3\times10^{-3},10^{-3},0.08)$. In this case, the surface density at the planet's orbit, $\Sigma(R_p)$, converges at $\sim10^3t_p$. Note that the corresponding viscous time for gap-opening according to Equation \ref{t_nu} is 
$\sim1.3\times10^3t_p$.

\subsubsection{Measurement of Radial Gap Structure}
To characterize the radial gap structure and quantify the gap depth, we introduce three measurements. First, we define the radial surface density profile, denoted as $\Sigma(R)$, which is obtained by azimuthally averaging the surface density while excluding the azimuth around the planet: 

\begin{equation}
\Sigma(R)=\frac{\int_{\delta\phi}^{2\pi-\delta\phi}\Sigma(R,\phi)d\phi}{\int_{\delta\phi}^{2\pi-\delta\phi}d\phi},
\end{equation}
We set $\delta\phi=\pi/12$. The azimuthal position of the planet is $\phi_p=0$.

Next, we determine the gap depth, denoted as $\Sigma_{\rm gap}$, which corresponds to the average surface density within a narrow ring centered at the planet's location. Specifically, we compute $\Sigma_{\rm gap}$ using the following expression:

\begin{equation}
\Sigma_{\rm gap}=\frac{\int_{R_p-\Delta}^{R_p+\Delta}dR\int_{\delta\phi}^{2\pi-\delta\phi}Rd\phi\Sigma(R,\phi)}{\int_{R_p-\Delta}^{R_p+\Delta}dR\int_{\delta\phi}^{2\pi-\delta\phi}Rd\phi},
\end{equation}
Here, $\Delta$ represents the half-width of the narrow ring, set to be $\text{max}(R_H,h_p)$, the maximum value between the Hill radius $R_H$ and the scale height at the planet's location $h_p$. This is half the value set in \citet{2014ApJ...782...88F}. We do so to obtain the gap depth closer to the center of the gap.

In addition, we determine the radial width of the gap by locating its inner ($R_{\rm in}$) and outer edges ($R_{\rm out}$) corresponding to the positions where the surface density reaches the geometric mean of the minimum and undepleted values \citep{2017ApJ...835..146D},
\begin{equation}
    \Sigma_{\rm edge}=\Sigma(R_{\rm in})=\Sigma(R_{\rm out})=
    \sqrt{\Sigma_0\Sigma_{\rm min}}.
    \label{def_width}
\end{equation}
The gap width is then 
\begin{equation}
    w_{\rm gap}=R_{\rm out}-R_{\rm in}.
\end{equation}

\subsubsection{Simulation Results}
\begin{figure*}[ht]
\centering
\begin{minipage}[b]{0.9\textwidth}
{\includegraphics[width=\textwidth]{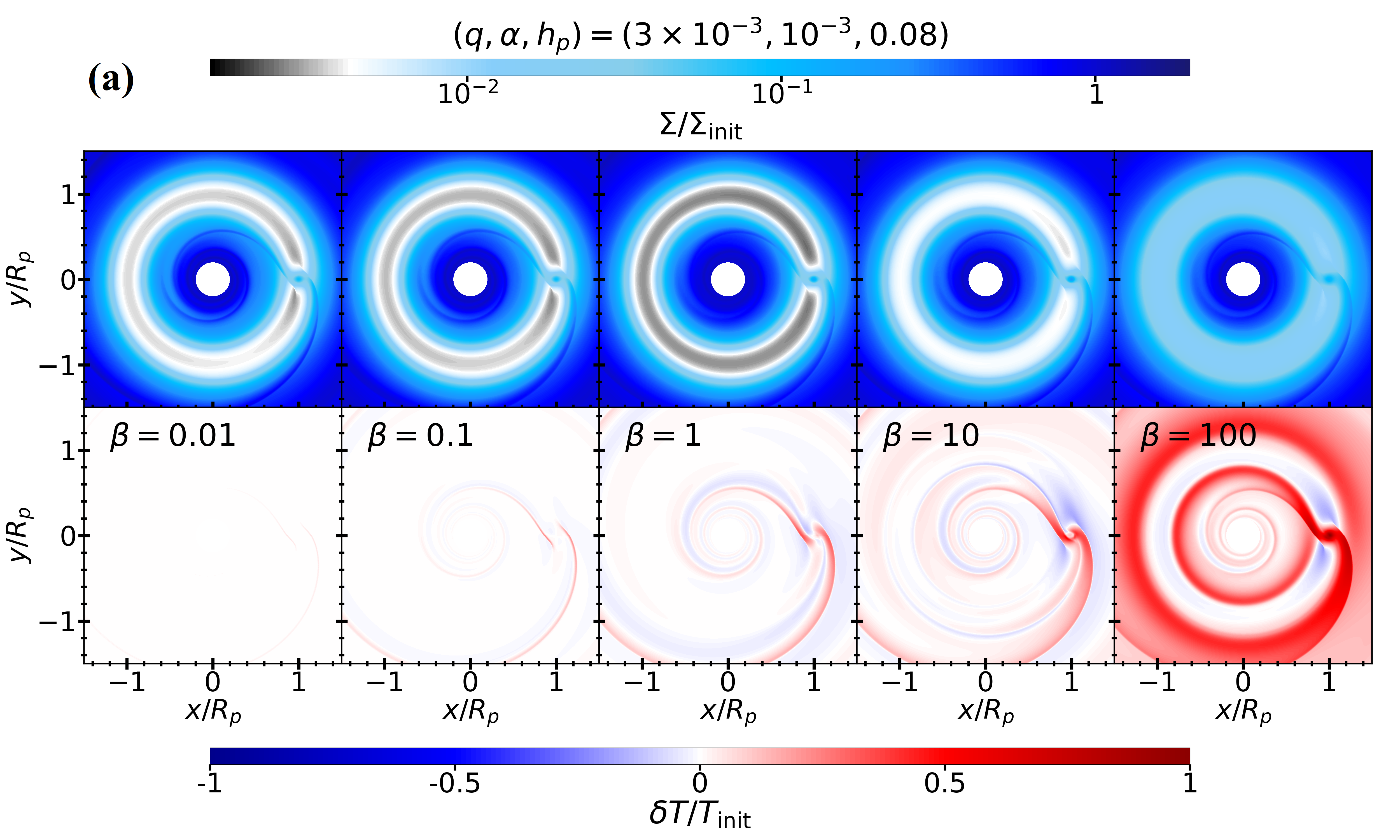}}
\end{minipage}
\begin{minipage}[b]{0.9\textwidth}
{\includegraphics[width=\textwidth]{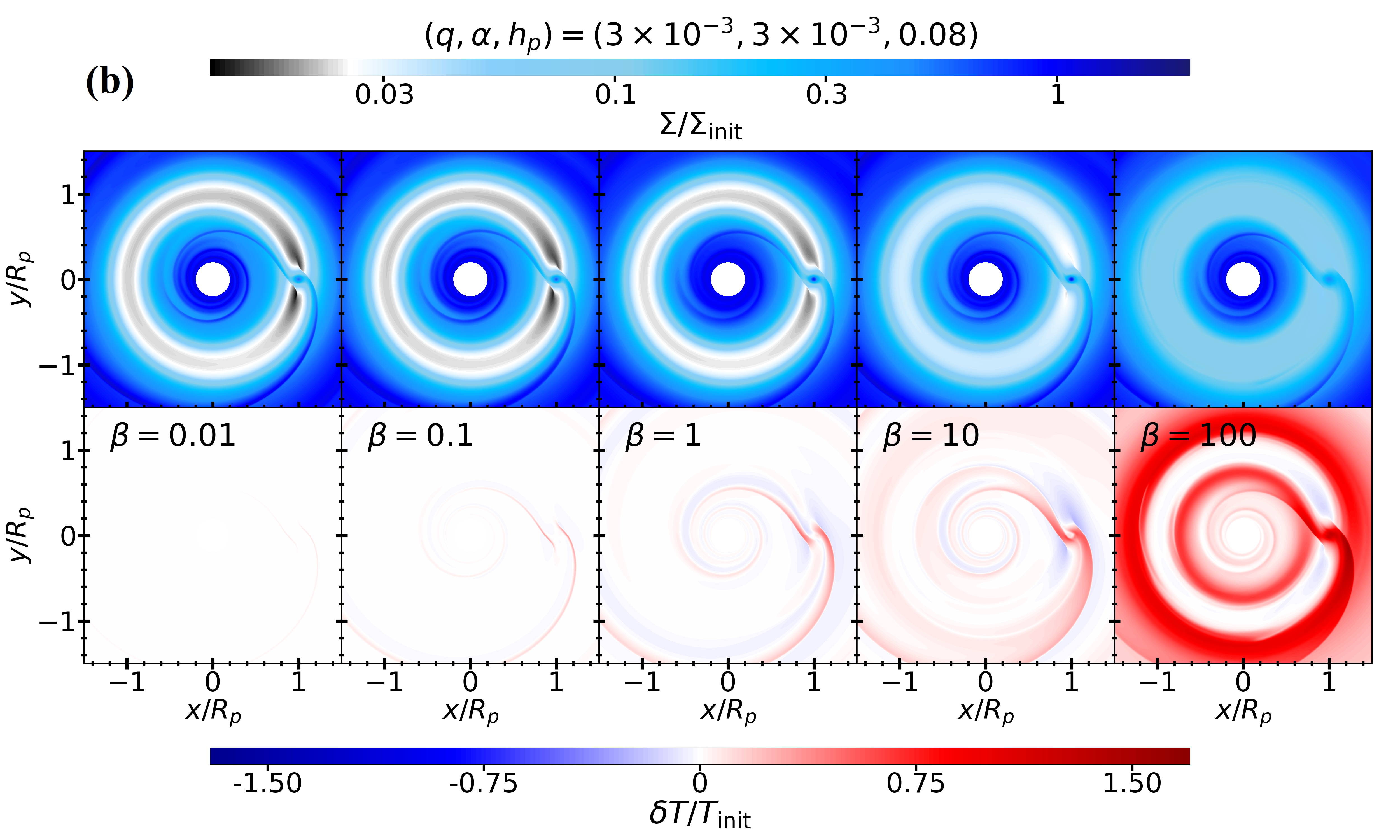}}
\end{minipage}
\caption{\label{simulation} 
The quasi-steady state gaseous density profile $\Sigma/\Sigma_{\rm init}$ and temperature perturbations $\delta T/T_{\rm init}$ for (a) a low viscosity case (Run h8-6) with parameters $(q,\alpha,h_p)=(3\times10^{-3},10^{-3},0.08)$, and (b) a high viscosity case (Run h8-7) with parameters $(q,\alpha,h_p)=(3\times10^{-3},3\times10^{-3},0.08)$.  
The dimensionless thermal relaxation timescale $\beta$ are $0.01,0.1,1,10$, and $100$, respectively, increasing from left to right.
The lower end of the color bar for surface density (top row in each case) is designed to highlight the fine difference at the gap center across different panels.
In low-viscosity disks, as $\beta$ increases, the gap reaches its maximum depth when the thermal relaxation timescale matches the dynamical timescale $(\beta=1)$, with no significant change in width. For relatively high $\alpha$ values $(\geq3\times10^{-3})$, the gap becomes shallower and wider monotonically with $\beta$ (especially when $\beta>1$).}
\end{figure*}

Figure \ref{simulation} presents the surface density structure in a steady-state configuration, along with the corresponding temperature perturbation relative to the cooling time scale $\beta$. 
Notably, in the upper panels of Figure \ref{simulation}, the low-viscosity case $(\alpha=10^{-3})$ exhibits a minimum gap depth at $\beta\approx1$. However, in the high-viscosity case $(\alpha=3\times10^{-3})$ in the bottom panels, the gap depth decreases monotonically with $\beta$.

\begin{figure*}[ht]
\centering
\includegraphics[width=\textwidth]{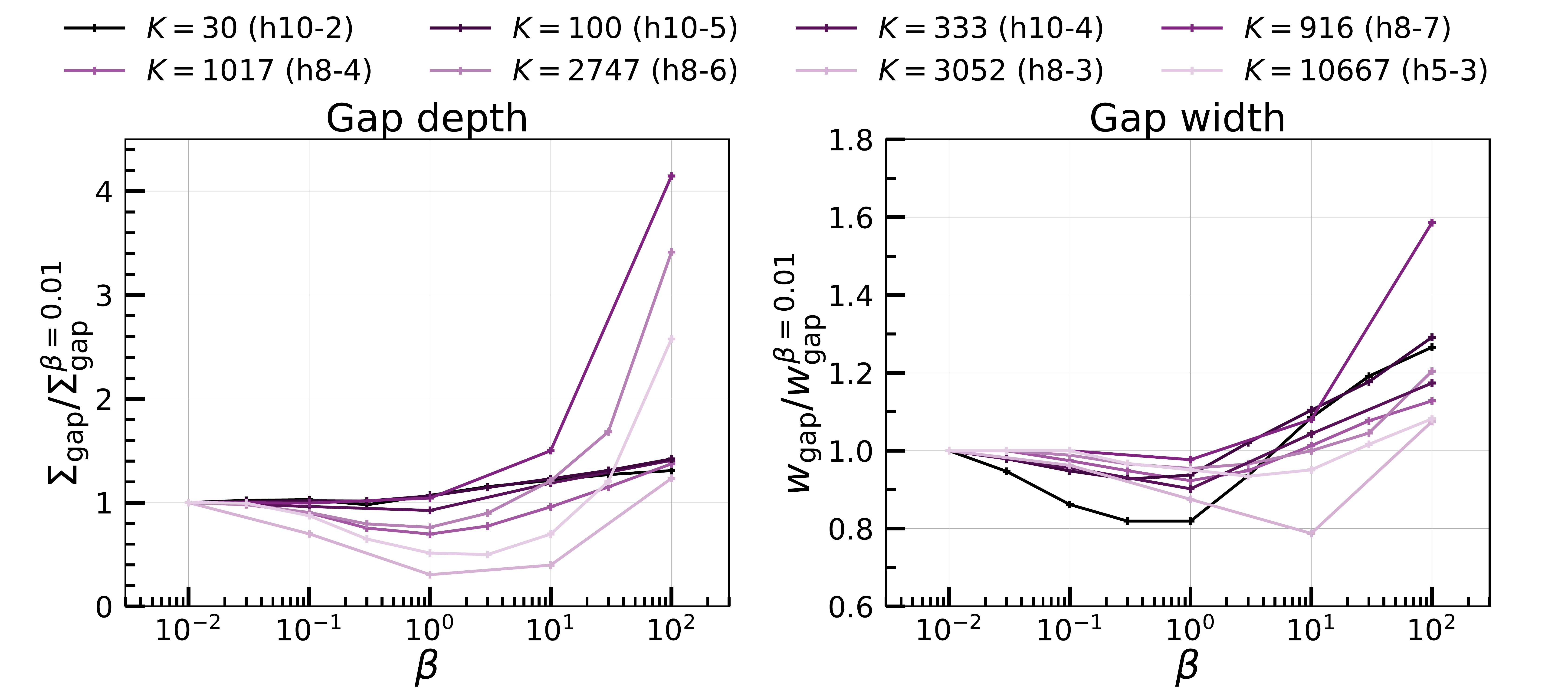}
\caption{\label{depth_width} The simulated gap depth (left panel) and width (right panel), both plotted versus the cooling timescale $\beta$. For each value of $K$, the gap depth and width are normalized with respect to the values obtained at $\beta=0.01$, approximating the isothermal case. 
The viscous parameter $\alpha$ varies from $10^{-4}$ to $3\times10^{-3}$. In the range of $K$ from approximately $10^2$ to $10^4$, we observed that the gap reaches its maximum depth and minimum width at around $\beta=1$. As $K$ increases, the change in gap depth becomes more pronounced, whereas the change in width becomes less significant. 
The data used in this figure is tabulated in Table \ref{table:depth_width}. 
}
\end{figure*}

Figure \ref{depth_width} shows the numerically-derived gap depth ($\Sigma_{\rm gap}$) and width ($w_{\rm gap}$) as a function of cooling timescale ($\beta$). The simulated gap depth and width with $K\sim10^1$ to $K\sim10^4$
are listed in Table \ref{table:depth_width}. 

\section{Empirical Models for Gap Depth and Width}
\label{sec:gapdepth}
In this section, we propose an empirical model that incorporates thermodynamic considerations to describe the gap depth. Drawing upon previous studies, we aim to generalize the gap depth formula to include the impact of the cooling time scale, $\beta$. To verify and provide insights into this model, we numerically investigate the time evolution of torques and angular momentum flux (AMF) with their dependence on $\beta$.

\subsection{Torque Balance and Angular Momentum Transfer}
The evolution of angular momentum carried by density waves plays a crucial role in shaping the disk. Therefore, understanding the dynamics of angular momentum flux (AMF) and torque balance is essential for comprehending the origins of different gap structures.

\begin{figure}[h]
\centering
\includegraphics[width=0.4\textwidth]{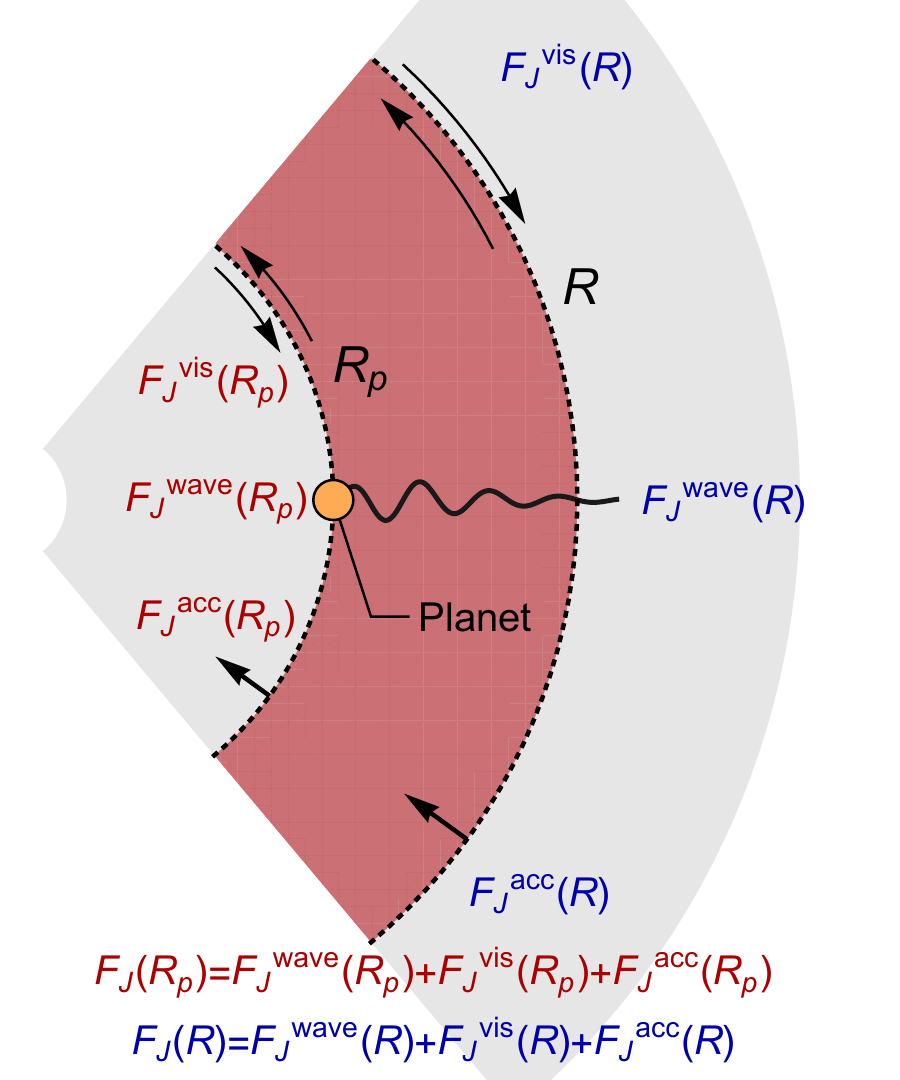}
\caption{\label{sketch_amf}  A sketch map depicting the components of the total AMF at an arbitrary radius $R$. The conservation of angular momentum is investigated within the red ring region (taking the outer disk for example). The total AMF flow out the region is $F_J(R)-F_J(R_p)=(F_J^{\rm vis}(R)+F_J^{\rm wave}(R)+F_J^{\rm acc}(R))-(F_J^{\rm vis}(R_p)+F_J^{\rm wave}(R_p)+F_J^{\rm acc}(R_p))$. 
} 
\end{figure}

Consider a region of the disk spanning from the planetary orbit at $R_p$ to an arbitrary radius $R$ near the gap's outer edge (see Figure \ref{sketch_amf} for details). The total AMF, denoted as $F_J$, can be divided into three distinct components \citep{2017PASJ...69...97K}:

(1) AMF resulting from accretion flow:
\begin{equation}
F_J^{\rm acc}(R) = Rv_\phi^{\rm avg}(R)(2\pi R\langle v_r\Sigma(R)\rangle_\phi),
\label{FJacc}
\end{equation}
where 
$2\pi R\langle v_r\Sigma(R)\rangle_\phi$
represents the accretion rate at radius $R$.

(2) AMF originating from viscous diffusion:
\begin{equation}
F_J^{\rm vis}(R)=2\pi R^3\langle\nu\Sigma\frac{\partial\Omega}{\partial R}\rangle_\phi+2\pi R\langle\nu\Sigma\frac{\partial v_R}{\partial\phi}\rangle_\phi,
\label{FJvis}
\end{equation}
where $\nu$ denotes the kinematic viscosity, $\Sigma$ represents the surface density, and $\Omega$ and $v_R$ denote the local angular velocity and radial velocity, respectively. The angled brackets $\langle\rangle$ denote azimuthal averaging.

(3) AMF arising from non-axisymmetric advection due to 
density waves
\citep{2020MNRAS.493.2287Z,2020ApJ...892...65M,2020ApJ...904..121M}:
\begin{equation}
F_J^{\rm wave}(R)=2\pi R^2\langle\Sigma(R,\phi)\delta v_\phi \delta v_R\rangle_\phi,
\label{FJwave}
\end{equation}
where $\delta v_\phi$ and $\delta v_R$ represent the deviations in azimuthal and radial velocities 
from initial condition,
respectively, caused by the presence of the planet. 

The process of gap formation can be understood as a competition between the planetary torque, which aims to open gaps, and viscous diffusion, which tends to close gaps \citep{2020apfs.book.....A}. Considering the region from $R_p$ to $R$ (the red region in Figure \ref{sketch_amf}), the conservation of angular momentum can be expressed as:
\begin{equation}
\frac{dL}{dt} = \Gamma_p(R) - F_J(R) + F_J(R_p),
\end{equation}
where $L$ represents the angular momentum of the gas in this region. The cumulative torque within this radius range is denoted as $\Gamma_p(R)$:
\begin{equation}
\Gamma_p(R) = \int_{R_p}^R 2\pi R dR \langle\Sigma(R,\phi)\frac{\partial\Phi}{\partial\phi}\rangle_\phi,
\label{Tp}
\end{equation}

The total angular momentum outflow from this region is given by $F_J(R) - F_J(R_p)$. In a steady state, angular momentum conservation dictates that $\frac{dL}{dt} = \Gamma_p(R) - F_J(R) + F_J(R_p) = 0$. \cite{2017PASJ...69...97K} pointed out that it is the cumulative planetary deposition torque, defined as:
\begin{equation}
\Gamma_d(R) = \Gamma_p(R) - F_J^{\rm wave}(R),
\label{Td}
\end{equation}
rather than 
$\Gamma_p(R)$, that governs the gap-opening process. The former represents the absorbed portion of the latter within the annular region $(R_p,R)$ of the disk.

To examine the dependence of gap depth on planetary mass and disk properties, we employ the conservation of angular momentum in the annulus extending from $R_p$, the radius of the planet, to $R_+$, the outer edge of the gap \citep{2015ApJ...806L..15K}. Once a steady state is achieved, the deposition torque in this annulus should balance the net angular momentum flux entering the region from the inner and outer boundaries, which includes the contributions from accretion flow $(F_J^{\rm acc})$ and viscous diffusion $(F_J^{\rm vis})$. This can be expressed as:
\begin{equation}\begin{aligned}
\Gamma_d(R) = &[-F_J^{\rm acc}(R) + F_J^{\rm vis}(R)]\\
        - &[-F_J^{\rm acc}(R_p) + F_J^{\rm vis}(R_p)].
\end{aligned}\end{equation}

A classic assumption, as proposed in \cite{2020apfs.book.....A}, is that the gap is wide enough to absorb all of the one-sided planetary torque ($\Gamma_d(R_+)\approx \Gamma_p(R_+)$), and it is narrow enough that $R_+ \approx R_p$ and $\Omega_K(R_+)\approx\Omega_p$. Under this assumption, the accretion-driven AMF entering and leaving the region cancels out ($F_J^{\rm acc}(R_+) - F_J^{\rm acc}(R_p)\approx 0$). The cumulative deposition torque balances the viscous diffusion at the inner $(R_p)$ and outer $(R_+)$ boundaries of the annular gap region. This can be written as:
\begin{equation}
\Gamma_d(R_+) = F_J^{\rm vis}(R_+) - F_J^{\rm vis}(R_p).
\label{Td2}
\end{equation}
Considering a disk rotating at Keplerian velocity $\Omega_K(R) = \sqrt{GM_*} R^{-3/2}$, we can approximate the viscous diffusion term $F_J^{\rm vis}(R)$ as:
\begin{equation}
F_J^{\rm vis}(R) \approx 3\pi\alpha (h_pR^{1/4})^2R^4\Omega_K^2\Sigma.
\label{FJvis_approx}
\end{equation}
By evaluating Equation (\ref{FJvis_approx}) at $R_+$ and $R_p$, we obtain $F_J^{\rm vis}(R_+)$ and $F_J^{\rm vis}(R_p)$ in Equation (\ref{Td2}):
\begin{equation}\begin{aligned}
F_J^{\rm vis}(R_+) &= 3\pi\alpha h_p^2\Sigma(R_+) R_+^4\Omega^2(R_+),\\
F_J^{\rm vis}(R_p) &= 3\pi\alpha h_p^2\Sigma(R_p) R_p^4\Omega_p^2
.\label{FJRp}
\end{aligned}\end{equation}

Note that the unperturbed surface density at $R_+$ is 
$\Sigma(R_+)=\Sigma_0(R_+/R_p)^{-1}$,
while the angular velocity is approximately $\Omega_K^2(R_+)\propto\frac{1}{R_+^3}$, and they cancel out with $R_+^4$. Then, 
the viscous AMF at $R_+$ becomes 
\begin{equation}\begin{aligned}
F_J^{\rm vis}(R_+) = 3\pi\alpha h_p^2\Sigma_0 R_p^4\Omega_p^2.\label{FJR+}
\end{aligned}\end{equation}
We introduce the one-sided Lindblad torque as an approximation for the cumulative deposition torque at $R_+$ \citep{2020apfs.book.....A}:
\begin{equation}\begin{aligned}
\Gamma_d(R_+) = f_0\Sigma(R_p)q^2h_p^{-3}R_p^4\Omega_p^2,
\label{Td_approx}
\end{aligned}\end{equation}
where $f_0 = 0.12\pi$ is a constant derived from the integration of the WKB torque formula \citep{1979MNRAS.186..799L}. Substituting Equations (\ref{FJRp}) and (\ref{Td_approx}) into Equation (\ref{Td2}), we obtain:
\begin{equation}\begin{aligned}
f_0\Sigma(R_p)&q^2h_p^{-3}R_p^4\Omega_p^2\\
& = 3\pi\alpha h_p^2\Sigma_0 R_p^4\Omega_p^2 - 3\pi\alpha h_p^2\Sigma(R_p) R_p^4\Omega_p^2,
\end{aligned}\end{equation}
which provides a more precise formula for the gap depth:
\begin{equation}\begin{aligned}
\frac{\Sigma_{\rm gap}}{\Sigma_0} = \frac{1}{1 + \frac{f_0}{3\pi}K},
\label{K}
\end{aligned}\end{equation}
where
\begin{equation}\begin{aligned}
K = q^2\alpha^{-1}h_p^{-5},
\end{aligned}\end{equation}
is a parameter that quantifies the planet's ability to open a gap. Numerical simulations conducted by \cite{2015ApJ...806L..15K}, \cite{2017PASJ...69...97K}, \cite{2014ApJ...782...88F}, and others have verified Equation \ref{K} as an accurate predictor of gap depth for given values of $(q, \alpha, h_p)$.

Because in our sample most planetary masses are relatively large, the excitation of nonlinear density waves by these planets poses a challenge. As a result, the gaps predicted by Eq. (\ref{K}) tend to underestimate the actual gaps observed in simulations. To address this issue, modifications to Eq. (\ref{K}) are necessary.

\subsection{Nonlinear effects}
When $K>10^3$, due to the nonlinearity of density waves excited by the planets, the gap will be much wider than in linear cases, so that Eq. (\ref{Td_approx}) will no longer be a good approximation for $\Gamma_d(R_+)$. As shown in Figure \ref{Td_beta}, the numerically simulated $\Gamma_d$ at the edge of the gap will be greater than the estimated value from Eq. (\ref{Td_approx}), and this effect becomes more pronounced as $K$ increases. As a result, if nonlinear effects are not taken into account, the predicted gap will appear shallower.

To tackle this problem, 
\cite{2017PASJ...69...97K} proposed a model that incorporates a nonlinear factor, obtained by multiplying the torque density, and employs an iterative algorithm to generate the gap profile. However, for the sake of simplicity in our fitting formula, we adopt an exponential modification $e^{\lambda K}$ for the term representing the planetary deposition torque ($\Gamma_d(R_+)$), which is recognized as the driving force for gap-opening at the outer edge of the gap. 
Eq. (\ref{K}) thus becomes
\begin{equation}\begin{aligned}
\frac{\Sigma_{\rm gap}}{\Sigma_0}=\frac{1}{1+\frac{f_0}{3\pi}e^{\lambda_0 K}K}.
\label{K_nl}
\end{aligned}\end{equation}
This modification allows us to incorporate the nonlinear effects when $K>10^3$.

\subsection{Formula for gap depth considering the cooling time scale}
\label{formula}
\begin{figure*}[ht]
\centering
\includegraphics[width=\textwidth]{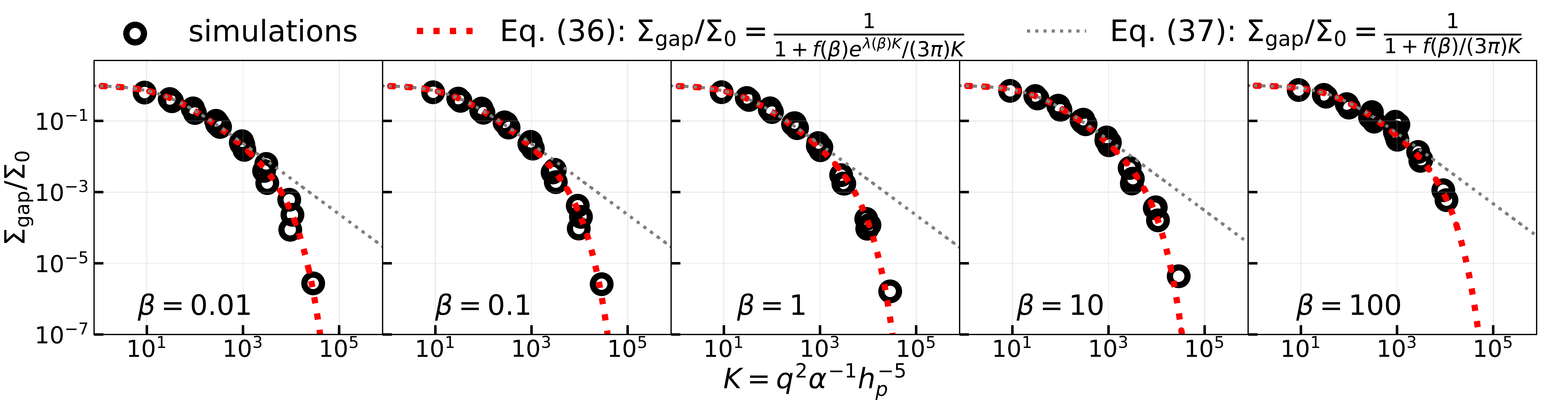}
\caption{\label{sigma_K} 
Averaged value of the azimuthally averaged surface density in the gap area $\Sigma_{\rm gap}$ versus the parameter $K$. The cooling time scales are 0.01,0.1,1,10,100, from left to right. Simulations are marked with circles, 
where different colors represent varying cooling time scales.
The red and gray dotted lines are fitted curves according to Equations (\ref{K3}) \& (\ref{KL})
without and with the correction factor for nonlinearity,
respectively. 
} 
\end{figure*}
\begin{figure}[ht]
\centering
\includegraphics[width=0.44\textwidth]{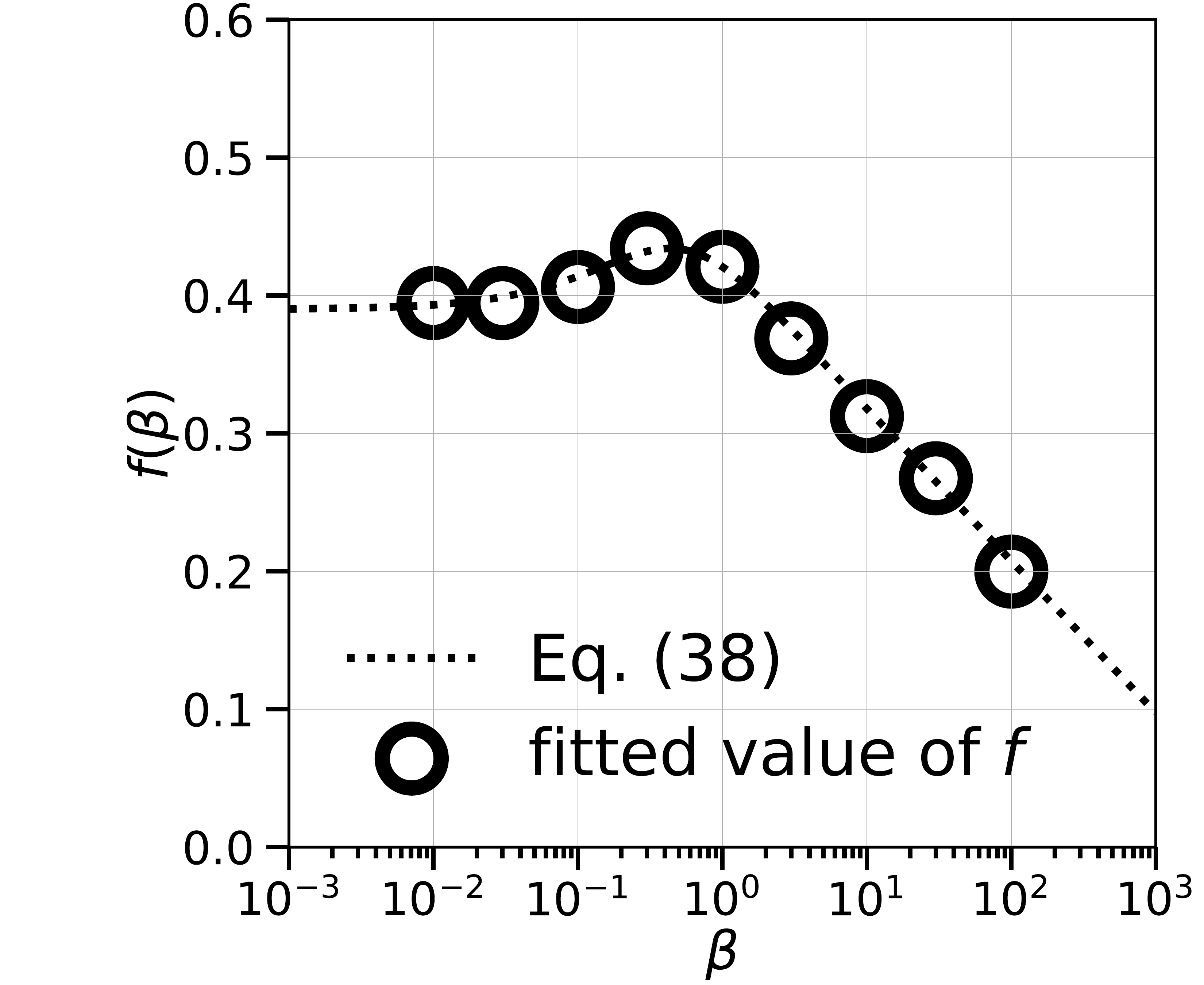}
\includegraphics[width=0.44\textwidth]{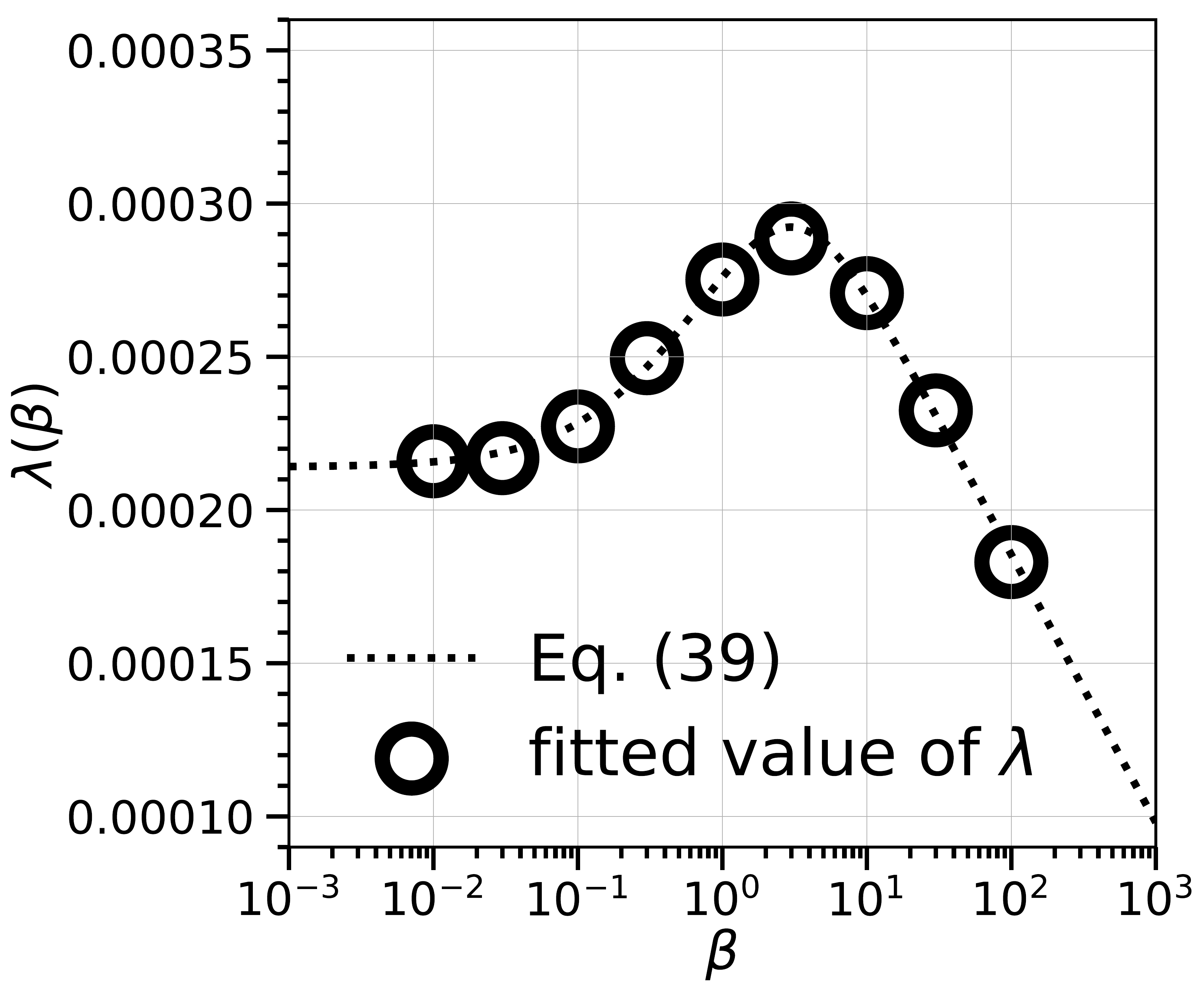}
\caption{\label{f_gamma}  
The dependence of the fitting parameters ($f(\beta)$, top, and $\lambda(\beta)$, bottom) in Eq. (\ref{K3}) as a function of $\beta$.
} 
\end{figure}
\begin{figure*}[ht]
\centering
\includegraphics[width=0.8\textwidth]{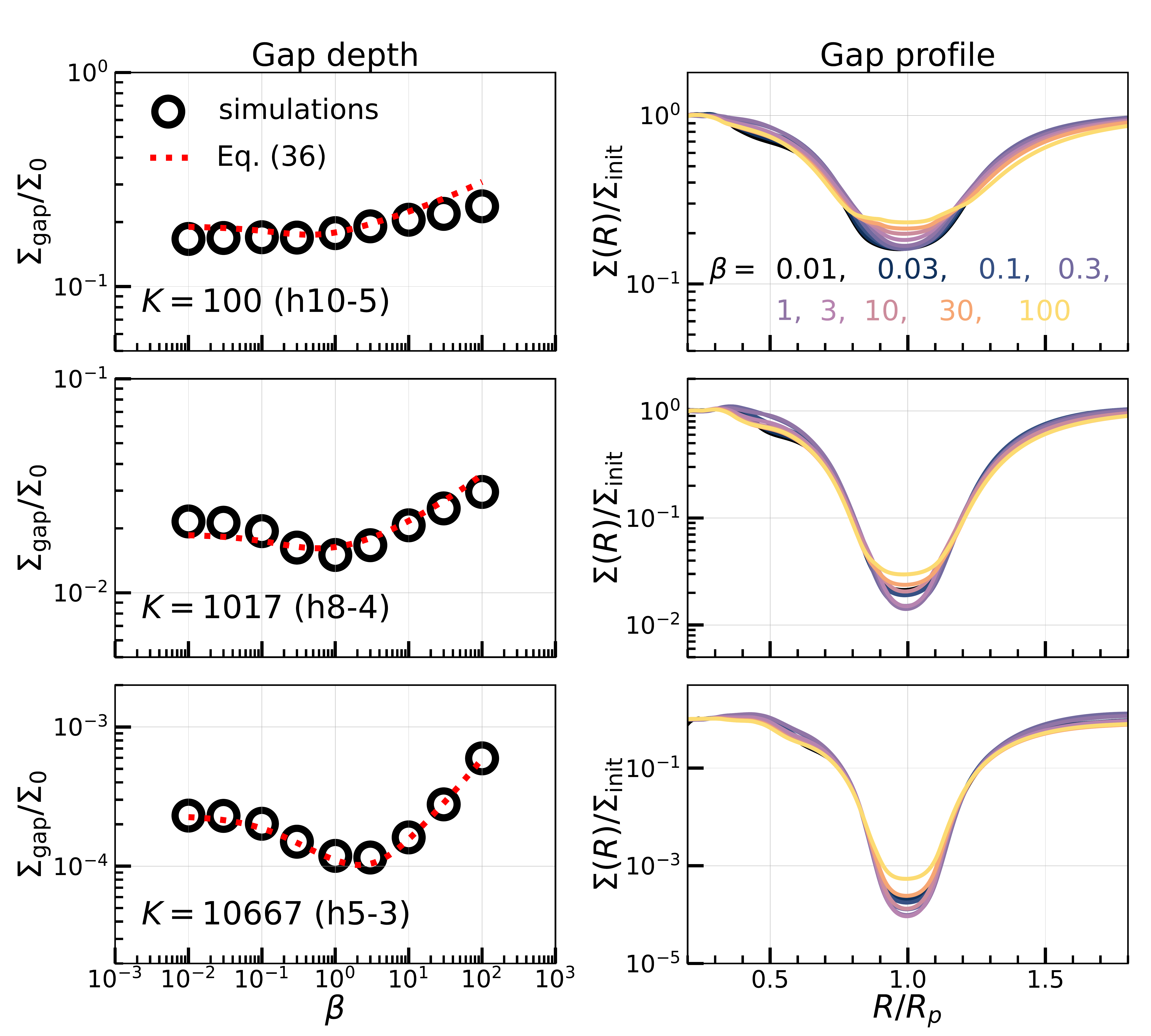}
\caption{\label{Fit_2} 
The dependence of gap depth (left) and gap profile (right) on the thermal relaxation timescale $\beta$ for different gap-opening parameters $K$. 
Left panel: gap depth $(\Sigma_{\rm gap}/\Sigma_0)$ as a function of cooling time scale $(\beta)$. 
Circles are simulation results. The black dotted curve is Eq. (\ref{K3}). Right panel: the azimuthally averaged gap profile with $\beta$ ranging from $0.01$ to $100$.} 
\end{figure*}
\begin{deluxetable}{lll}
\tablewidth{0.48\textwidth}
\tabcolsep=0.1cm
\tablecaption{Fitting formulas for gap depth\label{table:formula}}
\tablehead{
\colhead{Eq. number}&\colhead{formula}&\colhead{applicable range}
}
\startdata
Eq. (\ref{K})& $\frac{\Sigma_{\rm gap}}{\Sigma_0}=\frac{1}{1+\frac{f_0}{3\pi}K}$,& Linear, isothermal\\ 
Eq. (\ref{K_nl})& $\frac{\Sigma_{\rm gap}}{\Sigma_0}=\frac{1}{1+\frac{f_0}{3\pi}e^{\lambda_0K}K}$,&  Nonlinear, isothermal\\
Eq. (\ref{KL})&$\frac{\Sigma_{\rm gap}}{\Sigma_0}=\frac{1}{1+\frac{f(\beta)}{3\pi}K}$,&  Linear, with cooling\\
Eq. (\ref{K3})&$\frac{\Sigma_{\rm gap}}{\Sigma_0}=\frac{1}{1+\frac{f(\beta)}{3\pi}e^{\lambda(\beta)K}K}$,&  Nonlinear, with cooling
\enddata
\end{deluxetable}

In order to incorporate thermal effects, we introduce a dependence of 
the cooling timescale $\beta$
on $\lambda_0$ and $f_0$. The modified formula for the gap depth, considering the cooling time scale, is given by:
\begin{equation}\begin{aligned}
\frac{\Sigma_{\rm gap}}{\Sigma_0}=\frac{1}{1+\frac{f(\beta)}{3\pi}e^{\lambda(\beta)K}K}.
\label{K3}
\end{aligned}\end{equation}
Here, the constant coefficent $f_0=0.12\pi$ in Eq. (\ref{Td_approx}) corresponding to linear, isothermal cases is replaced by $f(\beta)e^{\lambda(\beta) K}$. 
This coefficient 
represents the ratio between the cumulative deposition torque $\Gamma_d(R_+)$ acting on the region from the planet's position to the periphery of the gap, and $F_{J0}=\Sigma(R_p)q^2h_p^{-3}R_p^4\Omega_p^2$. We utilize Equation (\ref{K3}) to fit the gap depth $\Sigma_{\text{gap}}$, obtained from approximately 100 numerical simulations considering the cooling effect.

In the pursuit of a fitting formula, we initially employ Equation \ref{K3} to fit the gap depth for a specific $\beta$ and derive the corresponding $f$ and $\lambda$ values.

In Figure \ref{sigma_K}, we present two fitted curves for gap depth: Equation (\ref{K3}), which incorporates a correction for the nonlinear effect, and
\begin{equation}
\frac{\Sigma_{\rm gap}}{\Sigma_0} = \frac{1}{1 + \frac{f(\beta)}{3\pi}K},
\label{KL}
\end{equation}
which shares the same form as Equation (\ref{K}) but replaces $f_0$ with $f(\beta)$, which depends on $\beta$. Figure \ref{sigma_K} illustrates that even when considering the cooling effect, Equation (\ref{K3}) continues to provide a satisfactory fit
but with a new coefficient $f(\beta)e^{\lambda(\beta)K}$ instead of $f_0$.
In addition, Equation (\ref{KL}) gives good performance when $K<10^3$.
The comparison of different deposition coefficient are shown in Fig. \ref{Td_beta}.

The fitted values of $f$ and $\lambda$ are marked by circles in Figure \ref{f_gamma}.
Subsequently, we adopt the functional form $f(\beta)=\text{log}_{10}\left(\frac{\beta^{-1}+c_1}{c_2\beta+\beta^{-1}}\right)^{\epsilon}+f_0$ to fit the coefficients $f$ and $\lambda$ as functions of the cooling timescale $\beta$. The results are presented in Equations (\ref{fit1}) and (\ref{fit2}).
\begin{equation}
f(\beta)=\text{log}_{10}\left(\frac{\beta^{-1}+7.0}{3.2\beta+\beta^{-1}}\right)^{0.11}+0.39
\label{fit1}
\end{equation}
\begin{equation}
\lambda(\beta)=\text{log}_{10}\left(\frac{\beta^{-1}+4.6}{0.099\beta+\beta^{-1}}\right)^{8.7\times10^{-5}}+2.1\times10^{-4}
\label{fit2}
\end{equation}
The deposition factor peaks at $\beta\approx1$, indicating that a cooling timescale comparable to the dynamical timescale enhances the absorption of excitation torque by the disk substance compared to the adiabatic and locally isothermal cases, as previously shown by \cite{2020ApJ...892...65M,2020ApJ...904..121M,2020MNRAS.493.2287Z,2023arXiv230514415Z}.

Figure \ref{Fit_2} illustrates the relationship between the gap depth and the cooling time scale $\beta$ for planets with varying gap-opening abilities. As the gap-opening capacity $K$ increases from $\sim10^1$ to $\sim 10^4$, the cooling time scale corresponding to the minimum value of $\Sigma_{\text{gap}}$ shifts from $0.3$ to $3$. 
We also compared the performance of the fitted gap depth curves between low-viscosity and high-viscosity disks. For slow cooling cases $(\beta\geq10)$ with a viscous parameter $\alpha\geq3\times10^{-3}$, the surface density in the gap may be underestimated due to viscous heating. 

All fitting formulas and their applicable ranges are summarized in Table \ref{table:formula}.
The torque balance (i.e., Eq. (\ref{Td2})) revealed in the gap depth formula Eq. (\ref{K3}) are numerically verified. $\Gamma_p$, $\Gamma_d$, $F_J^{\rm wave}$, and $F_J^{\rm vis}$ are performed according to Equations (\ref{Tp}), (\ref{Td}), (\ref{FJwave}), and (\ref{FJvis}), respectively. (See Appendix \ref{appendix} for detail.)

\subsection{Gap Width}
\begin{figure*}[ht]
\centering
\includegraphics[width=\textwidth]{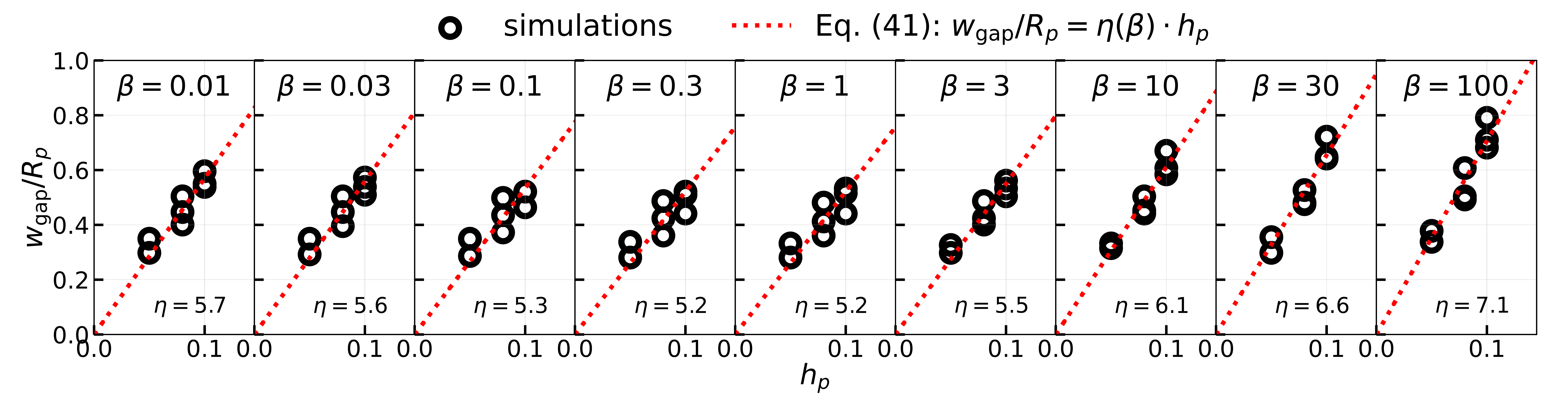}
\caption{
The gap width $w_{\rm gap}$ (defined by Eq. (\ref{def_width})) versus the disk aspect ratio $h_p$.
The cooling time scales are 0.01, 0.03, 0.1, 0.3, 1, 3, 10, 30, 100, from left to right. Simulations are marked with circles, while the dashed lines are the fitted linear relation between the two quantities (Eq. (\ref{width_formula})), whose slope ($\eta$) is labeled at the bottom. 
\label{width_hp} 
} 
\end{figure*}
\begin{figure}[ht]
\centering
\includegraphics[width=0.43\textwidth]{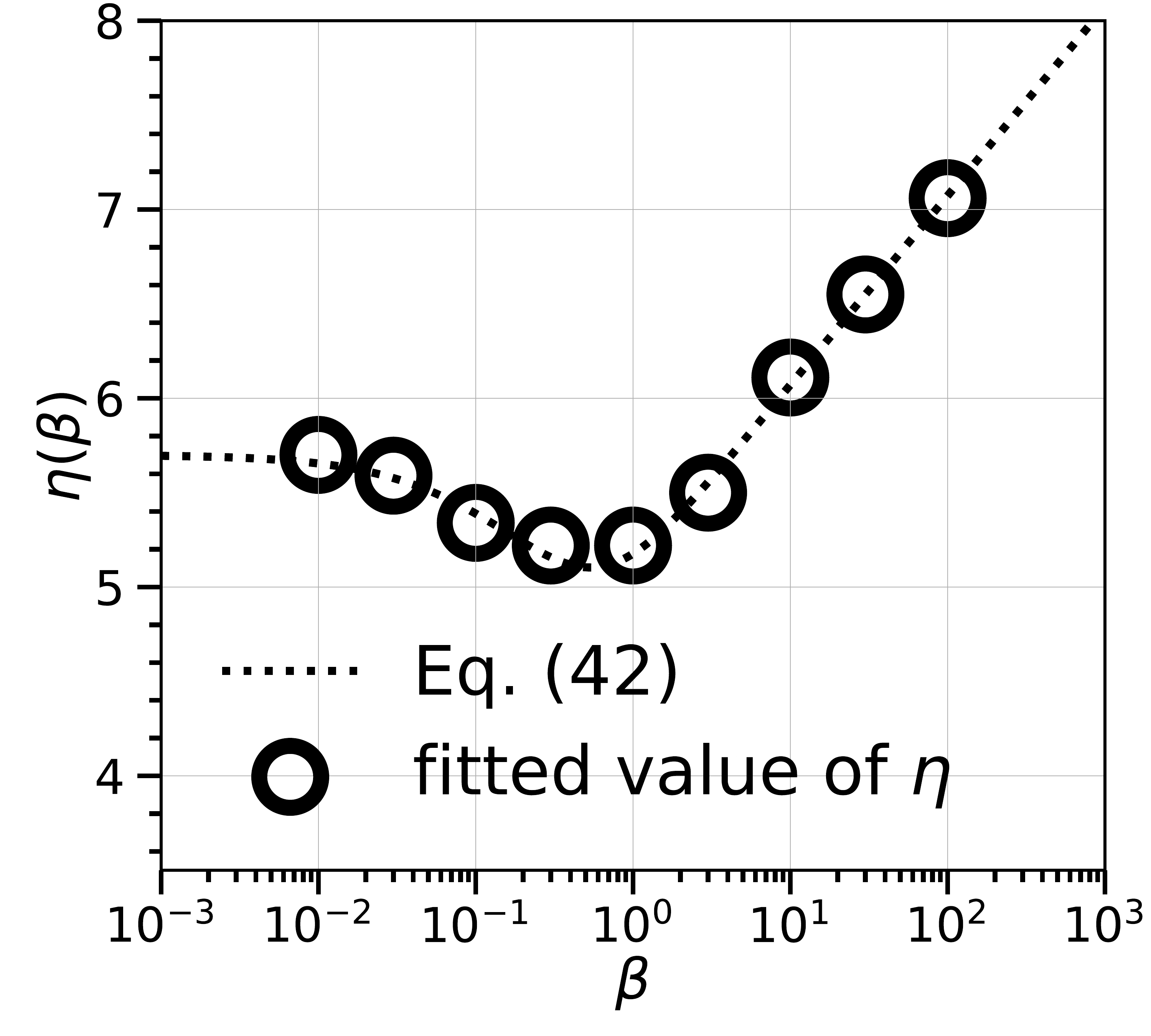}
\caption{
The dependence of $\eta$ (i.e., the slope of the fitting lines in Figure \ref{width_hp}) on $\beta$ (Eq. (\ref{eta_fit})).
\label{eta_fig} 
} 
\end{figure}
\begin{figure}[ht]
\centering
\includegraphics[width=0.4\textwidth]{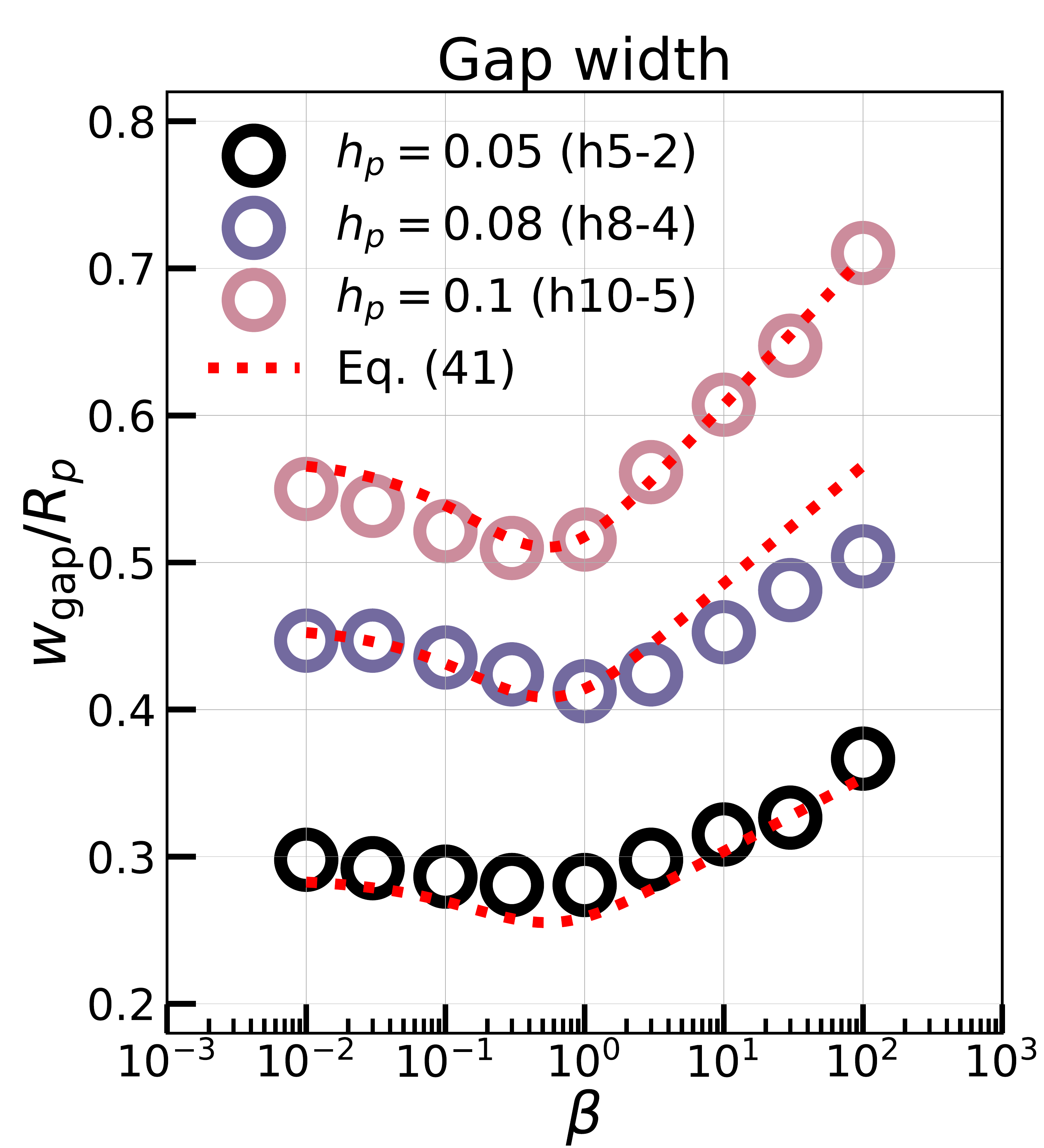}
\caption{
The dependence of gap width on $\beta$ for three disk aspect ratios. 
The red dotted curves are our fitting functions (Eq. (\ref{width_formula})).
\label{width_beta} 
} 
\end{figure}

An empirical formula for the dependence of gap width on $\beta$ is provided. 
Using isothermal simulations, \citet{2017ApJ...835..146D} found
\begin{equation}
w_{\rm gap}/R_p=\eta\cdot h_p,
\label{width_formula_iso}
\end{equation}
where $\eta$ is approximately a constant, 5.8.  
To take into account the cooling effect, we retain the format of the fitting function, but treat $\eta$ as a function of $\beta$, i.e.,
\begin{equation}
     w_{\rm gap}/R_p=\eta(\beta)\cdot h_p.
\label{width_formula}
\end{equation}
Figure \ref{width_hp} shows the fitted $\eta$ values at a variety of $\beta$. 
Finally, $\eta(\beta)$ is fitted as
\begin{equation}
   \eta(\beta)=-\text{log}_{10}\left(\frac{\beta^{-1}+11}{2.6\beta+\beta^{-1}}\right)+5.7,
\label{eta_fit}
\end{equation}
and shown in Figure \ref{eta_fig}.
Note that it has the same functional form as in $f$ (Eq. \ref{fit1}) and $\lambda$ (Eq. \ref{fit2}). 
The comparison between the gap width predicted by the fitting formula and numerical simulations is presented in Figure \ref{width_beta} for three different $h_p$.


\section{Discussion}
\label{sec:discussion}
\subsection{Applications to real PPDs}
By incorporating a thermal relaxation term during our numerical simulations, we have generalized the results regarding the gap depth and width for massive planets under the locally isothermal condition \citep{2014ApJ...782...88F,2015ApJ...806L..15K,2016PASJ...68...43K,2017PASJ...69...97K}. We provide approximate analytical expressions for the gap depth and width in terms of $\alpha$, $q$, $h_p$, and $\beta$, which accurately describe the gap depth and width within the range $10^{-4} \leq \alpha \leq3\times10^{-3}$, $3\times10^{-4} \leq q\leq3\times10^{-3}$, $0.05 \leq h_p \leq 0.1$, and $10^{-2}\leq\beta\leq10^{2}$.

Our empirical formulas for gap depth and width carries implications for observations. 
In recent years, numerous gap and cavity structures have been discovered by ALMA, not only in dust but also in gas \citep{2016A&A...585A..58V,Yen_2016,2016PhRvL.117y1101I}. These observational measurements, when combined with theoretical models, allow for the constraints of various properties, such as planetary mass, for the planets residing within the disk.

For example, utilizing spatially resolved CO isotopologue emission observed by ALMA, \cite{2015A&A...579A.106V,2016A&A...585A..58V,2017ApJ...836..201D} 
derived gas surface density distributions and identified significant drops up to several orders of magnitude in gas surface density inside central cavities in a few disks. This discovery implies the presence of close-in Jovian planets.
While recent observations have achieved order of magnitude accuracy in characterizing the gap or cavity depth in gas, our study reveals that the gap depth can differ by roughly an order of magnitude when considering different thermal relaxation time scales. This finding highlights the significant impact of the radiative cooling in theoretical models on constraining planetary mass.

The depth and width of CO gaps in five PPDs were reported by \cite{Zhang_2021} from the ALMA large program Molecules with ALMA at Planet-forming Scales (MAPS), among which three (AS 209, HD 163296, and MWC 480) were reported to exhibit significant CO gaps. Gaussian functions were used in \cite{Zhang_2021} to fit the CO gap profiles and measure the gap depth and width in these three PPDs. In the remaining two PPDs, GM Aur has a cavity, while IM Lup has a very wide inner gap
(see Figure 18 in \cite{Zhang_2021} for details.) 
In addition, an HCO+ gas gap in HL Tau was found in \cite{2019ApJ...880...69Y}.

\begin{figure*}[ht]
\centering
\includegraphics[width=\textwidth]{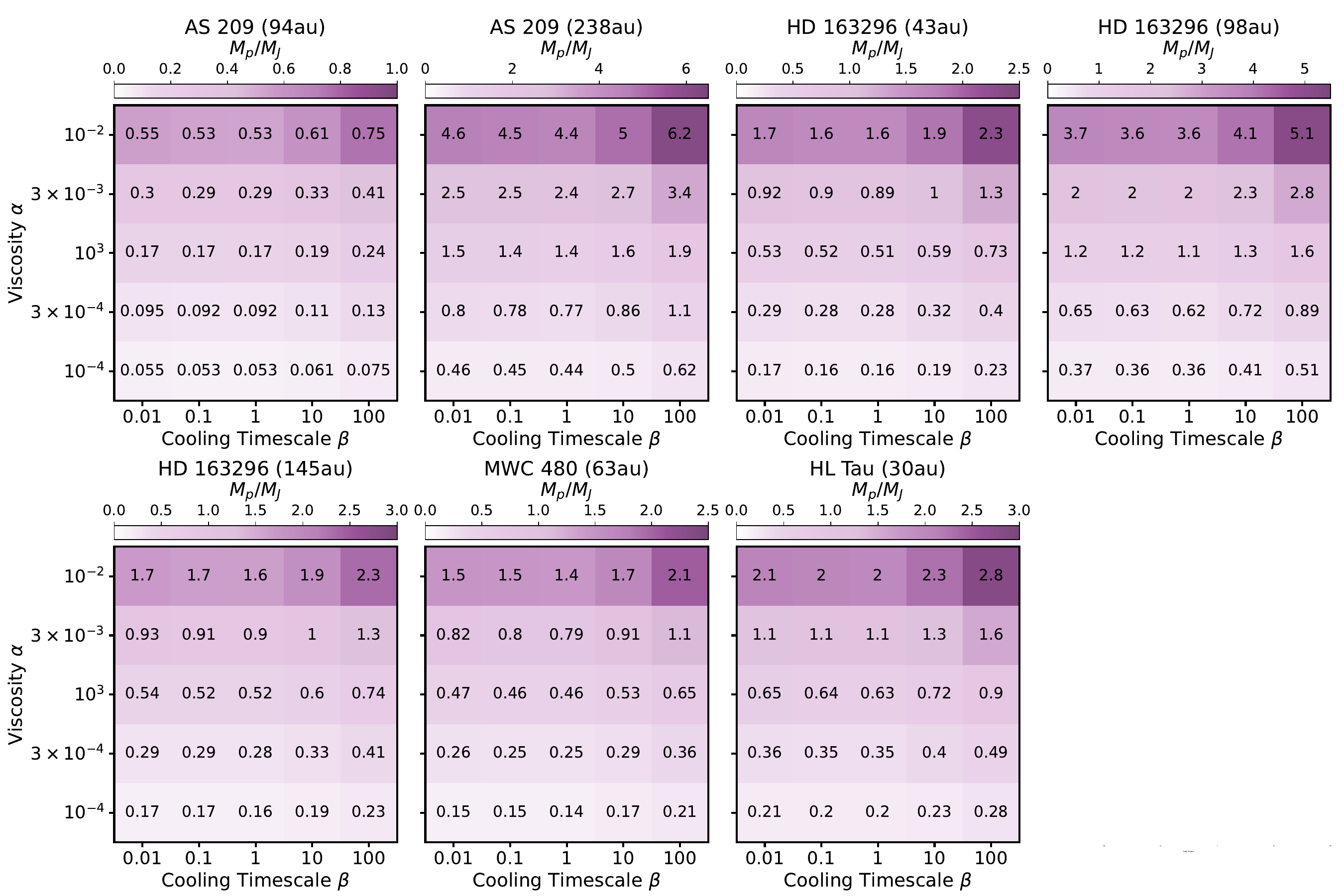}
\caption{
Expected planet mass ($M_p/M_J$) responsible for opening the gas gaps listed in Table \ref{tab:app}.  
The planetary mass is determined using Eq. (\ref{K3}) and 
inferred  $\Sigma_{\rm gap}/\Sigma_0$ and $h_p$ from observations, with viscosity ($\alpha$) ranging from $10^{-4}$ to $10^{-2}$ and cooling time scale ($\beta$) ranging from $10^{-2}$ to $10^2$. 
\label{planetmass} 
} 
\end{figure*}

Under the assumption that each gap is opened by one planet, we applied our fitting formula to the gas gaps mentioned above  
(see Table \ref{tab:app}, Column (1)). 
We derive the expected planetary mass for $5\times5=25$ combinations of $\alpha$ and $\beta$, shown in Figure \ref{planetmass}, based on Eq. \ref{K3} with observed gap depth and disk scale height in Table \ref{tab:app}. 
Our $\alpha$ ranges from $10^{-2}$ to $10^{-4}$, and $\beta$ ranges from 0.01 to 100.

\begin{figure}[ht]
\centering
\includegraphics[width=0.49\textwidth]{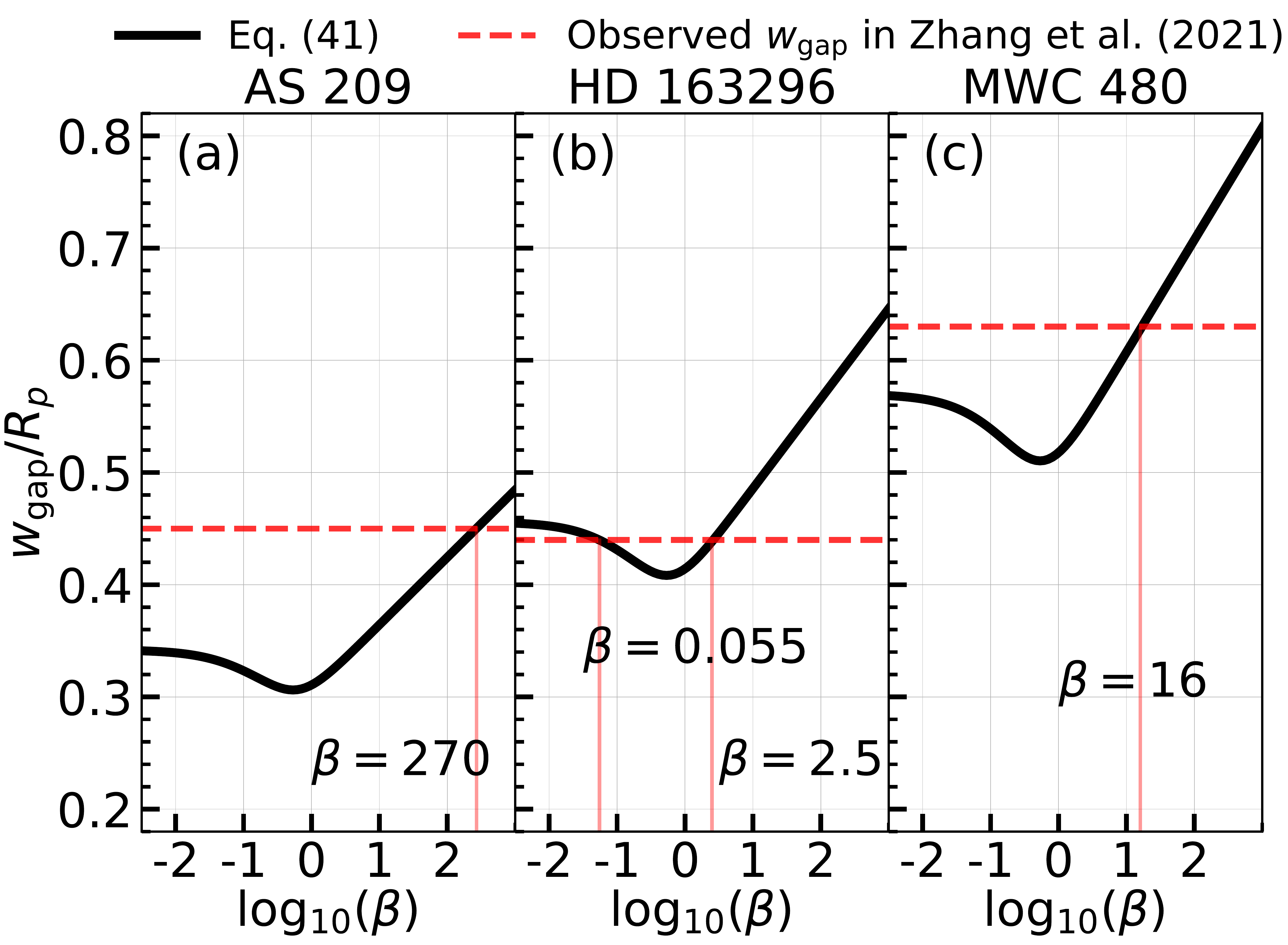}
\caption{
Predicted values of $\beta$ for three PPDs according to Eq. (\ref{width_formula}) and observed gap width ($w_{\rm gap}$) of (a) the 94 au gap of AS 209, (b) the 98 au gap of HD 163296, and (c) the 63 au gap of MWC 480, provided in Table \ref{tab:app}. 
\label{predict_beta} 
} 
\end{figure}

Furthermore, the fitting formula for gap width (i.e., Eq. \ref{width_formula}) can be used to constrain $\beta$. Among the observed gas gaps listed in Table \ref{tab:app}, we found it possible to estimate 
$\beta$ in 3 gaps in AS 209, HD 163296, and MWC 480 
by finding the intersection between the observed $w_{\rm gap}$ (dashed line in Fig. \ref{predict_beta}) and Eq. \ref{width_formula} (solid curve in Fig. \ref{predict_beta}). For the remaining gaps, our assumption may not be valid due to insufficient data to accurately measure the gap width.

It is crucial to emphasize that the observed gap properties pertain to CO and HCO$^+$ gaps, and whether they differ from those of H$_2$ gaps remains uncertain.
Besides, our estimates should be regarded as indicative values, considering that real disks may deviate from the assumptions in our models, 
e.g., uniform viscosity and cooling timescale across the gap region.

Accounting for cooling in observations, a basic consequence is that the isothermal models (e.g., \citet{2014ApJ...782...88F,2015ApJ...807L..11D,2017PASJ...69...97K}, etc.) lead to overestimation of a planet's mass when cooling timescale is around the dynamical timescale ($t_{\rm cool}\approx t_{\rm dyn}$) and underestimation when it exceeds $10t_{\rm dyn}$. 
For instance, in a disk with $(\alpha,h_p)=(10^{-3},0.05)$, the isothermal model predicts a $1M_J$ mass planet responsible for a deep gap with a $1.9\times10^{-3}$ drop of gaseous surface density. 
However, our thermal relaxation model suggests that a gap with the same depth will be created by a $1.3M_J$ planet with $\beta=100$, or a $0.9M_J$ planet with $\beta=1$.

Additionally, when considering cooling, the temperature distribution of the disk deviates from the locally isothermal scenario due to radiative cooling, viscous heating, and other factors. \cite{Muley_2021} demonstrates that temperature structures in spiral arms can serve as an additional theoretical reference for observations. The temperature profile across a gap (e.g., Fig. \ref{simulation}) may also encode information on the thermodynamics in the disk.


\begin{deluxetable}{lcccc}
\tabletypesize{\scriptsize}
\tablewidth{0.8\textwidth}                                     
\tabcolsep=0.1cm
\tablecaption{
Applying Our Empirical Formulas to Real PPDs\label{tab:app}
}
\tablehead{
\colhead{Source} & \colhead{$M_\star/M_\odot$}&  \colhead{$\Sigma_{\rm gap}/\Sigma_0$} & \colhead{$w_{\rm gap}/R_{\rm gap }$}&\colhead{$(H/R)_{\rm gap}$}\\
\colhead{(1)}& 
\colhead{(2)}& 
\colhead{(3)}&
\colhead{(4)}&
\colhead{(5)}
} 
\startdata
\multicolumn{5}{c}{CO gaps reported in \cite{Zhang_2021}}\\
\hline
AS~209 (94au)   &1.2&0.45&0.45&0.059  \\
AS~209 (236au)  &1.2&0.03&\nodata&0.074  \\
HD~163296 (43au)&2.0&0.51&\nodata&0.079  \\
HD~163296 (98au)&2.0&0.58&0.44&0.084  \\
HD~163296 (145au)&2.0&0.68&\nodata&0.087  \\
MWC~480 (63au)  &2.1&0.38&0.63&0.096   \\
\hline
\multicolumn{5}{c}{HCO$^+$ gap reported in \cite{2019ApJ...880...69Y}}\\
\hline
HL Tau (30au)   &2.1&0.22&\nodata&0.065 
\enddata
\tablecomments{
Colume(1): source name and gap position. 
Colume(2): stellar mass. 
Colume(3): Observed gas gap depth.
Colume(4): Observed gas gap width defined by the ratio between the FWHM and the location of the gap. 
Colume(5): disk aspect ratio at the gap location.
}
\tablecomments{The errors of the observed quantities are not considered in our applications.}
\end{deluxetable}

\subsection{Limitations and outlook}
Firstly, we have simplified our simulations by focusing solely on the evolution of gas distribution and neglecting the influence of dust feedback on the gas. Previous studies have shown differences in gaseous structures due to dust feedback, especially for wider and deeper gaps \citep{2019ApJ...879L..19K}. 
Disk self-gravity is another factor that could affect the evolution of the angular momentum flux (AMF) \citep{2020MNRAS.493.2287Z}. 
Therefore, an important future direction based on our research is to explore the effects of disk self-gravity and radiative cooling on gas-dust interactions.

Besides, we adopted a simple assumption of constant-$\beta$ thermal relaxation for the sake of parametric simplicity. In reality, $\beta$ may be a function of location in disks \citep{2020MNRAS.493.2287Z,muley2023threetemperature}.

On top of that, our $\beta$-cooling models, although convenient for parameter exploration, has limitations when applied to ALMA disks. Recent studies have highlighted the strong influence of in-plane cooling on spiral wave AMF and gap opening induced by planets \citep{2020ApJ...904..121M}. After that, \cite{2023arXiv230514415Z} found discrepancies between fully radiative models with direct cooling and models considering only surface cooling. 
By setting $\beta$ as an effective cooling timescale that incorporates both surface and in-plane cooling (e.g., \citealt{2020ApJ...904..121M}), our model is also able to take in-plane cooling into consideration, which could give a rough correction and make the results of AMFs, torques, and gap depth similar to those of 
fully radiative models with direct cooling term.
However, even with this modification, the output of our adjusted model will still differ from those with realistic cooling terms. 
As direct constraints on cooling timescales in observations are still limited (e.g., \citealt{Teague_2022}), we defer detailed studies in the effect of anisotropic cooling on gap properties to the future.

\section{Summary}
Through extensive 2D hydrodynamical simulations, we have explored the effect of radiative cooling on gas structures within protoplanetary disks
(see Figure \ref{simulation} for a direct visualization). 
We have quantified the 
dependence of gap depth and width on the cooling timescale $\beta$
(Eq. \ref{K3}, Fig. \ref{depth_width}, and Table \ref{table:depth_width}),
a dimensionless quantity normalized by the dynamical timescale 
$(t_{\rm dyn}=\Omega^{-1})$,
for planets with masses ranging from Saturn to 3 Jupiter masses, roughly equivalent to 0.6 to 6 disk thermal masses in a disk with a local aspect ratio $h_p= 0.08$. 
The gap is narrowest and deepest at $\beta \approx 1$, as found in \citet{2020ApJ...892...65M,2020ApJ...904..121M} and \citet{2020MNRAS.493.2287Z}, and widest and shallowest at $\beta = 100$.
Significant variations in the gap depth, up to an order of magnitude, were observed for different $\beta$ values. 
Additionally, a deviation of $10\%\sim20\%$ in width is observed compared to the isothermal case.
A line chart comparing the depth and width of gaps in disks with $\beta$ cooling and those in nearly locally isothermal disks are shown in Figure \ref{depth_width}.
A table listing the simulated gap depth and width for a wide range of gap-opening capacity can be found in Table \ref{table:depth_width}.

Our numerical calculations have allowed us to evaluate the radial distribution of angular momentum fluxes and torques, confirming the torque balance achieved within the system as it reaches a quasi-steady state
(see Figure \ref{AMF_t}). 
We have also discovered that the discrepancy in steady-state gap depth corresponding to different cooling time scales can be attributed to 
the varying ability of density waves in propagation
(Figure \ref{AMF1}). 
Notably, around the critical value of $\beta=1$, the density wave experiences the strongest damping closest to the planet's position, leading to a more rapid decay of its carried angular momentum flux with increasing radius. This suggests that when the cooling timescale is comparable to the dynamical timescale, the disk material shows the highest absorption efficiency for the planetary excitation torque, as opposed to cases of immediate cooling (isothermal limit) and slow cooling (adiabatic limit). As a result, this leads to the formation of the deepest and narrowest gap.

In addition,
by fitting the surface density near the planet's orbit obtained from our numerical simulations, we have generalized 
the formula for gap depth, Eq. (\ref{K}),  
derived from torque balance for a steady-state gap 
in the linear and isothermal case. 
By introducing a correction factor $e^{\lambda K}$ for nonlinear effects and 
another correction factor $f(\beta)$ to account for the effect of cooling (Table \ref{table:formula}),
we have formulated an analytical expression 
(Eq. (\ref{K3}))
that effectively captures the relationship between gap depth and $(q, \alpha, h, \beta)$ in both linear and nonlinear regimes.
Gap depth as a function of $\beta$ is shown in Figs. \ref{Fit_2}. This is an extension of previous works concerning the dependence of gap depth on $(q, \alpha, h)$ alone (\citealt{2014ApJ...782...88F}; \citealt{2017PASJ...69...97K}).

An empirical formula for gap width with cooling effect (Eq. \ref{width_formula}), as an extension to the isothermal case reported in \cite{Dong_2017}, is also provided. The gap edges are defined as the locations where the surface density reaches the geometric mean of the minimum inside the gap and the undisturbed value. Gap width as a function of $\beta$ is shown in Fig. \ref{width_beta}.

We applied our empirical model (Eq. \ref{K3} \& \ref{width_formula}) to the observations of 6 CO gaps reported in the ALMA large program MAPS by \cite{Zhang_2021} and 1 HCO$^+$ gap observed in \cite{2019ApJ...880...69Y}. We inferred the masses of the possible planets responsible for opening these gaps (Figure \ref{planetmass}) and estimated the cooling time scale for 3 of them (Figure \ref{predict_beta}).

\bgroup
\section*{Acknowledgements}
We thank the anonymous referee for the thoughtful and constructive questions and suggestions. M.Z. appreciates Cong Yu for his insightful guidance on the semi-analytical models for planet-induced gaps. We also extends thanks to Shangfei Liu, Shunquan Huang, Jeffrey Fung, Shangjia Zhang, and Xilei Sun for helpful discussions.
P.H. and R.D. are supported by the Natural Sciences and Engineering Research Council of Canada (NSERC), the Alfred P. Sloan Foundation, and the Government of Canada's New Frontiers in Research Fund (NFRF), [NFRFE-2022-00159].
This research was enabled in part by support provided by Cedar and the Digital Research Alliance of Canada \url{alliance.can.ca}. 
We appreciate the Chinese Center for Advanced Science and Technology for hosting the Protoplanetary Disk and Planet Formation Summer School in 2022, organized by Xue-ning Bai and Ruobing Dong, which facilitated this research.
\egroup

\appendix
\section{Data of gap depth and width in quasi-steady states}
The simulated gap depth and width with various $(q,\alpha,h_p,\beta)$ from numerical simulations are listed in Table \ref{table:depth_width}.
\begin{deluxetable}{lcccccl}
\tabletypesize{\scriptsize}
\tabcolsep=0.3cm
\tablecaption{Simulated gap depth and width\label{table:depth_width}}
\tablewidth{0.9\textwidth}
\tablehead{
\nocolhead{a} & \multicolumn5c{Gap depth$(\Sigma_{\rm gap}/\Sigma_0)$/Gap width$(w_{\rm gap}/R_p)$} & \nocolhead{g} \\
\cline{2-6}
\colhead{$(q,\alpha,h_p)$} & \colhead{$\beta=0.01$} & \colhead{$\beta=0.1$} & \colhead{$\beta=1$} & \colhead{$\beta=10$} & \colhead{$\beta=100$} & \colhead{$K$} 
}
\startdata
$(3\times10^{-4},1\times10^{-3},0.1)$   & $6.3\times10^{-1}/0.60$ & $6.4\times10^{-1}/0.52$ & $6.4\times10^{-1}/0.53$ & $7.0\times10^{-1}/0.67$ & $7.4\times10^{-1}/0.79$ &  9    \\
$(3\times10^{-4},3\times10^{-4},0.1)$   & $4.1\times10^{-1}/0.53$ & $4.2\times10^{-1}/0.46$ & $4.4\times10^{-1}/0.43$ & $5.0\times10^{-1}/0.59$ & $5.4\times10^{-1}/0.68$ &  30   \\
$(1\times10^{-3},3\times10^{-3},0.1)$   & $3.6\times10^{-1}/0.56$ & $3.7\times10^{-1}/0.53$ & $3.9\times10^{-1}/0.54$ & $4.3\times10^{-1}/0.66$ & $4.8\times10^{-1}/0.88$ &  33   \\
$(3\times10^{-4},1\times10^{-4},0.1)$   & $2.3\times10^{-1}/0.52$ & $2.1\times10^{-1}/0.46$ & $2.4\times10^{-1}/0.40$ & $3.0\times10^{-1}/0.52$ & $3.0\times10^{-1}/0.64$ &  90   \\
$(3\times10^{-4},3\times10^{-4},0.08)$  & $2.3\times10^{-1}/0.40$ & $2.2\times10^{-1}/0.37$ & $2.3\times10^{-1}/0.36$ & $2.6\times10^{-1}/0.44$ & $2.9\times10^{-1}/0.49$ &  92   \\
$(1\times10^{-3},1\times10^{-3},0.1)$   & $1.7\times10^{-1}/0.55$ & $1.7\times10^{-1}/0.52$ & $1.8\times10^{-1}/0.52$ & $2.1\times10^{-1}/0.61$ & $2.4\times10^{-1}/0.71$ &  100  \\
$(3\times10^{-4},1\times10^{-4},0.08)$  & $1.1\times10^{-1}/0.41$ & $9.9\times10^{-2}/0.37$ & $9.9\times10^{-2}/0.37$ & $1.2\times10^{-1}/0.41$ & $1.3\times10^{-1}/0.48$ &  275  \\
$(3\times10^{-4},1\times10^{-3},0.05)$  & $8.0\times10^{-2}/0.30$ & $8.5\times10^{-2}/0.29$ & $8.3\times10^{-2}/0.28$ & $9.1\times10^{-2}/0.32$ & $1.2\times10^{-1}/0.37$ &  288  \\
$(1\times10^{-3},3\times10^{-4},0.1)$   & $6.8\times10^{-2}/0.53$ & $6.5\times10^{-2}/0.50$ & $6.3\times10^{-2}/0.48$ & $8.1\times10^{-2}/0.55$ & $9.5\times10^{-2}/0.62$ &  333  \\
$(3\times10^{-3},1\times10^{-3},0.1)$   & $2.5\times10^{-2}/0.57$ & $2.3\times10^{-2}/0.56$ & $2.0\times10^{-2}/0.54$ & $2.8\times10^{-2}/0.58$ & $4.9\times10^{-2}/0.69$ &  900  \\
$(3\times10^{-3},3\times10^{-3},0.08)$  & $2.3\times10^{-2}/0.50$ & $2.3\times10^{-2}/0.50$ & $2.4\times10^{-2}/0.49$ & $3.4\times10^{-2}/0.54$ & $9.5\times10^{-2}/0.79$ &  916 \\
$(3\times10^{-4},3\times10^{-4},0.05)$  & $2.7\times10^{-2}/0.29$ & $2.6\times10^{-2}/0.28$ & $2.1\times10^{-2}/0.27$ & $2.7\times10^{-2}/0.29$ & $3.8\times10^{-2}/0.32$ &  960  \\
$(1\times10^{-3},3\times10^{-4},0.08)$  & $2.2\times10^{-2}/0.45$ & $2.0\times10^{-2}/0.44$ & $1.6\times10^{-2}/0.41$ & $2.2\times10^{-2}/0.45$ & $3.1\times10^{-2}/0.50$ &  1017 \\
$(1\times10^{-3},3\times10^{-3},0.05)$  & $1.5\times10^{-2}/0.36$ & $1.7\times10^{-2}/0.36$ & $1.8\times10^{-2}/0.35$ & $2.5\times10^{-2}/0.39$ & $7.9\times10^{-2}/0.57$ &  1067 \\
$(3\times10^{-3},1\times10^{-3},0.08)$  & $3.9\times10^{-3}/0.50$ & $3.6\times10^{-3}/0.50$ & $3.0\times10^{-3}/0.48$ & $4.8\times10^{-3}/0.50$ & $1.3\times10^{-2}/0.61$ &  2747 \\
$(3\times10^{-4},1\times10^{-4},0.05)$  & $6.5\times10^{-3}/0.46$ & $4.6\times10^{-3}/0.44$ & $2.1\times10^{-3}/0.40$ & $2.9\times10^{-3}/0.36$ & $8.0\times10^{-3}/0.49$ &  3052 \\
$(1\times10^{-3},1\times10^{-3},0.05)$  & $1.9\times10^{-3}/0.36$ & $2.0\times10^{-3}/0.34$ & $1.8\times10^{-3}/0.34$ & $2.5\times10^{-3}/0.36$ & $8.6\times10^{-3}/0.44$ &  3200 \\
$(3\times10^{-3},3\times10^{-4},0.08)$  & $6.1\times10^{-4}/0.51$ & $4.1\times10^{-4}/0.51$ & $1.8\times10^{-4}/0.48$ & $3.5\times10^{-4}/0.48$ & $1.1\times10^{-3}/0.54$ &  9155 \\
$(1\times10^{-3},3\times10^{-4},0.05)$  & $2.3\times10^{-4}/0.35$ & $2.0\times10^{-4}/0.35$ & $1.2\times10^{-4}/0.33$ & $1.6\times10^{-4}/0.33$ & $5.9\times10^{-4}/0.38$ &  10667\\
$(3\times10^{-3},1\times10^{-3},0.05)$  & $2.9\times10^{-6}/0.40$ & $2.7\times10^{-6}/0.40$ & $1.7\times10^{-6}/0.39$ & $4.6\times10^{-6}/0.40$ & \nodata &  28800\\
\enddata
\end{deluxetable}

\section{Numerical analysis on torques and AMF}
\label{appendix}
The azimuthally averaged AMFs and torques are numerically computed to provide validation for Eq. \ref{K3}, the analytical formula for gap depth.

\subsection{Time evolution of AMFs and torques}
\begin{figure*}[h]
\centering
{\includegraphics[width=0.9\textwidth]{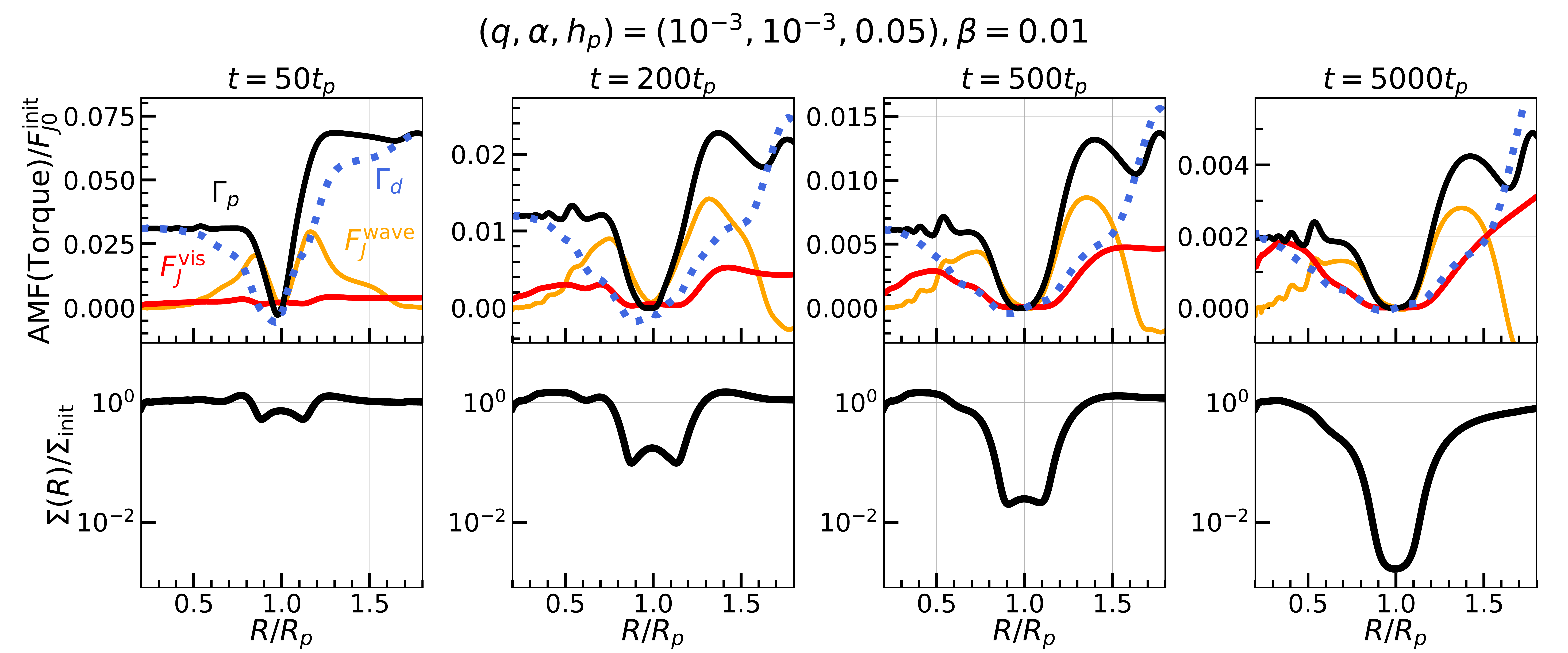}}
{\includegraphics[width=0.9\textwidth]{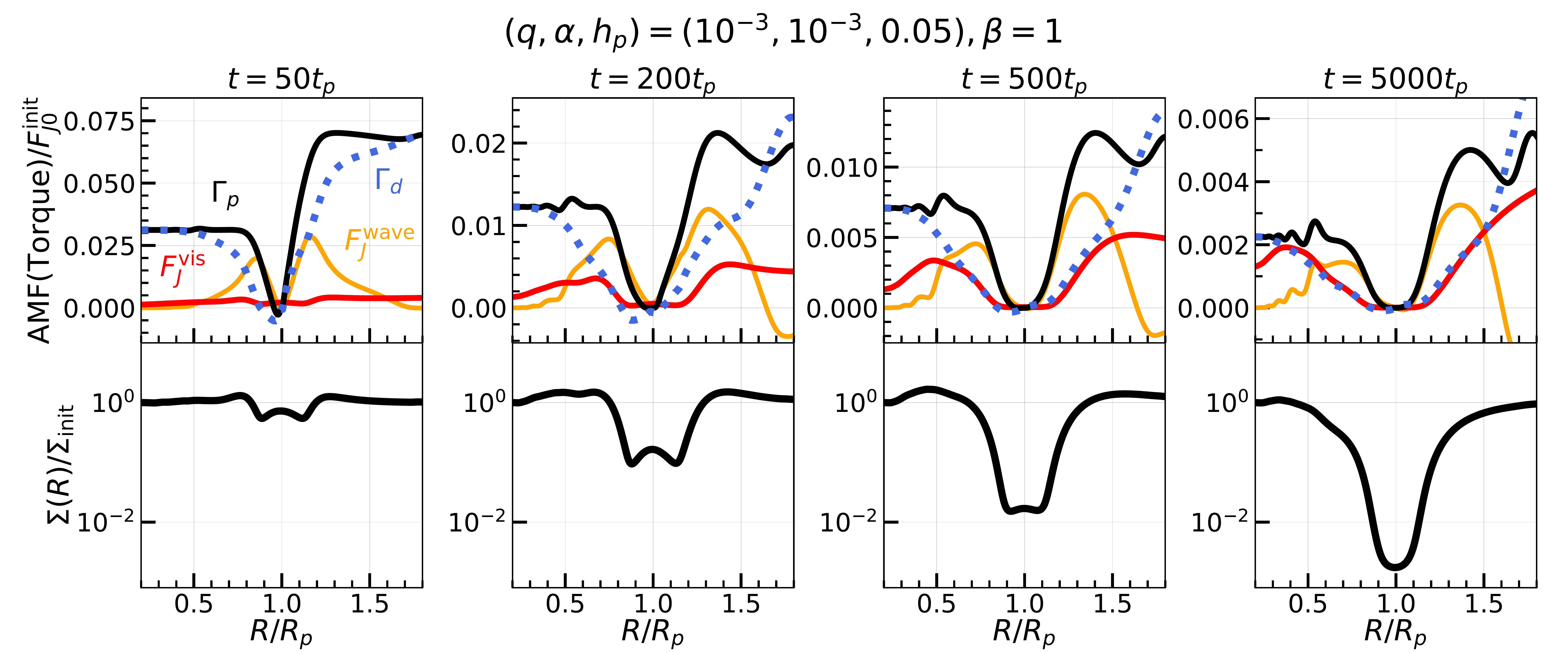}}
{\includegraphics[width=0.9\textwidth]{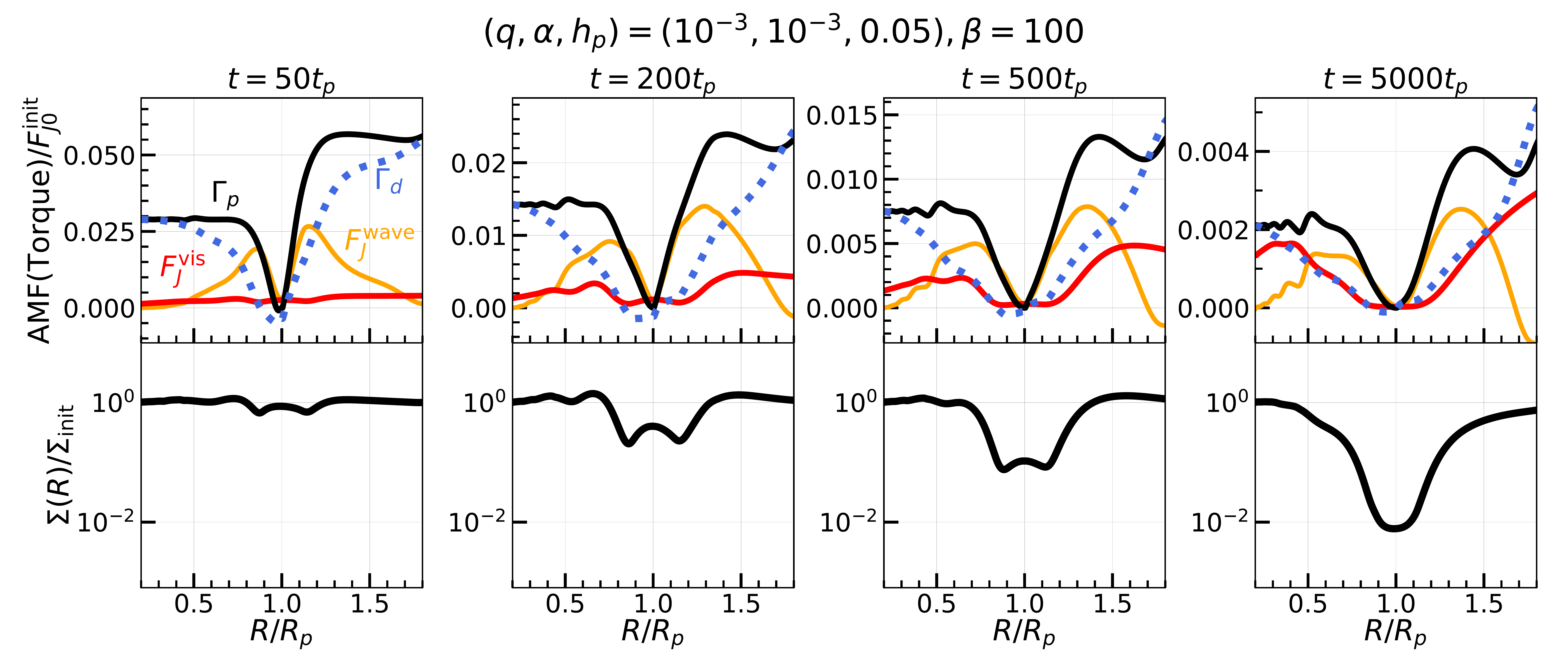}}
\caption{\label{AMF_t} 
The time evolution of AMF \& torque with $(q,\alpha,h_p)=(10^{-3},10^{-3},0.05)$. Four snapshots show the radial distribution of AMF and torque at $50$ orbits, $200$ orbits, $500$ orbits, and $5000$ orbits, respectively. The AMFs and torques are normalized by $F_{J0}^{\rm init}=\Sigma_0q^2h_p^{-3}R_p^4\Omega_p^2$. The top, central, and bottom panels correspond to three different cooling timescales $\beta$.}
\end{figure*}

We first investigate the temporal evolution of torque and the response of viscous diffusion to the gap structure induced by the gravitational disturbance. 
Figure \ref{AMF_t} presents four snapshots illustrating the radial profiles of AMF and torques during gap-opening process
in models with $(q, \alpha, h_p)=(10^{-3},10^{-3},0.05)$ and $\beta=0.01,1,100$. 
At $t=50$ orbits, the planetary torque $\Gamma_p$ significantly exceeds the viscous diffusion $F_J^{\text{vis}}$. 
During the gap-opening period ($t=100$ to $2000$ orbits), $\Sigma(R_p)$ experiences a relatively rapid decrease. Consequently, both the planet's excitation torque $\Gamma_p$ and the AMF due to density wave $F_J^{\text{wave}}$ decrease as well.
Meanwhile, the radial profile of viscous torque $F_J^{\text{vis}}$ undergoes only minor adjustments as a result of changes in surface density.

Over time, the deposition torque $\Gamma_d$ decreases to the same magnitude as the viscous torque $F_J^{\text{vis}}$, and the region of overlap between deposition torque and viscous torque gradually expands outward from the planetary orbit. This expansion continues until approximately $5000$ orbits when the deposition torque and viscous torque of the planet balance within a radial region of approximately $0.5R_p$ on either side of the planetary orbit. 

The right column of Figure \ref{AMF1} shows the AMF and torques in a quasi-steady state after evolving for $\sim5000$ orbits. The angular momentum stored in the ring region from $R_p$ to any arbitrary radius $R$ is precisely balanced by viscous diffusion at the boundaries $R$ and $R_p$. This indicates that the system approaches a steady state and torque balance prevails throughout the gap. Moreover, by comparing the simulated gap depth with the theoretical predictions, we observe that the gap depth remains constant thereafter.

As illustrated in Figure \ref{AMF_t}, when the system reaches a steady state, the profile of viscous AMF ($F_J^{\text{vis}}(R)$) closely resembles the cumulative deposited torque ($\Gamma_d(R)$). This suggests that in a protoplanetary disk with a stable gaseous gap formed by a planet, the density wave excited by the planet's gravity locally deposits a portion of its angular momentum into the disk material. The angular momentum gained by the ring region from $R_p$ to $R$ is promptly dissipated at the ring's boundary through viscosity. The remaining angular momentum continues to propagate with the density wave until it is fully damped. For a relatively deep gap ($\Sigma(R_p)/\Sigma_0<<1$), the viscous diffusion near the gap ($F_J^{\text{vis}}(R_p)\sim3\pi\nu\Sigma(R_p)\Omega_pR_p$) can be neglected compared to $F_J^{\text{vis}}(R_+)\sim3\pi\nu\Sigma_0\Omega_pR_p$.

\subsection{The dependence of quasi-steady-state AMFs and torques on $\beta$}\label{sec:AMF-beta}
We investigate the steady-state radial profiles of the planetary excitation torque ($\Gamma_p$), deposition torque ($\Gamma_d$), and AMF due to density wave ($F_J^{\text{wave}}$) to better understand the empirical formula in Equation \ref{K3}.

As mentioned in section \ref{formula}, the deposition coefficient represents the ratio between the theoretical value of the deposition torque $\Gamma_d(R_+)$ and the normalized flux $F_{J0}$, i.e.,
\begin{equation}
    \Gamma_d(R_+)/F_{J0}=f(\beta)e^{\lambda(\beta)K}.
\end{equation}

\begin{figure*}[h]
\centering
\subfigure[]{\includegraphics[width=0.45\textwidth]{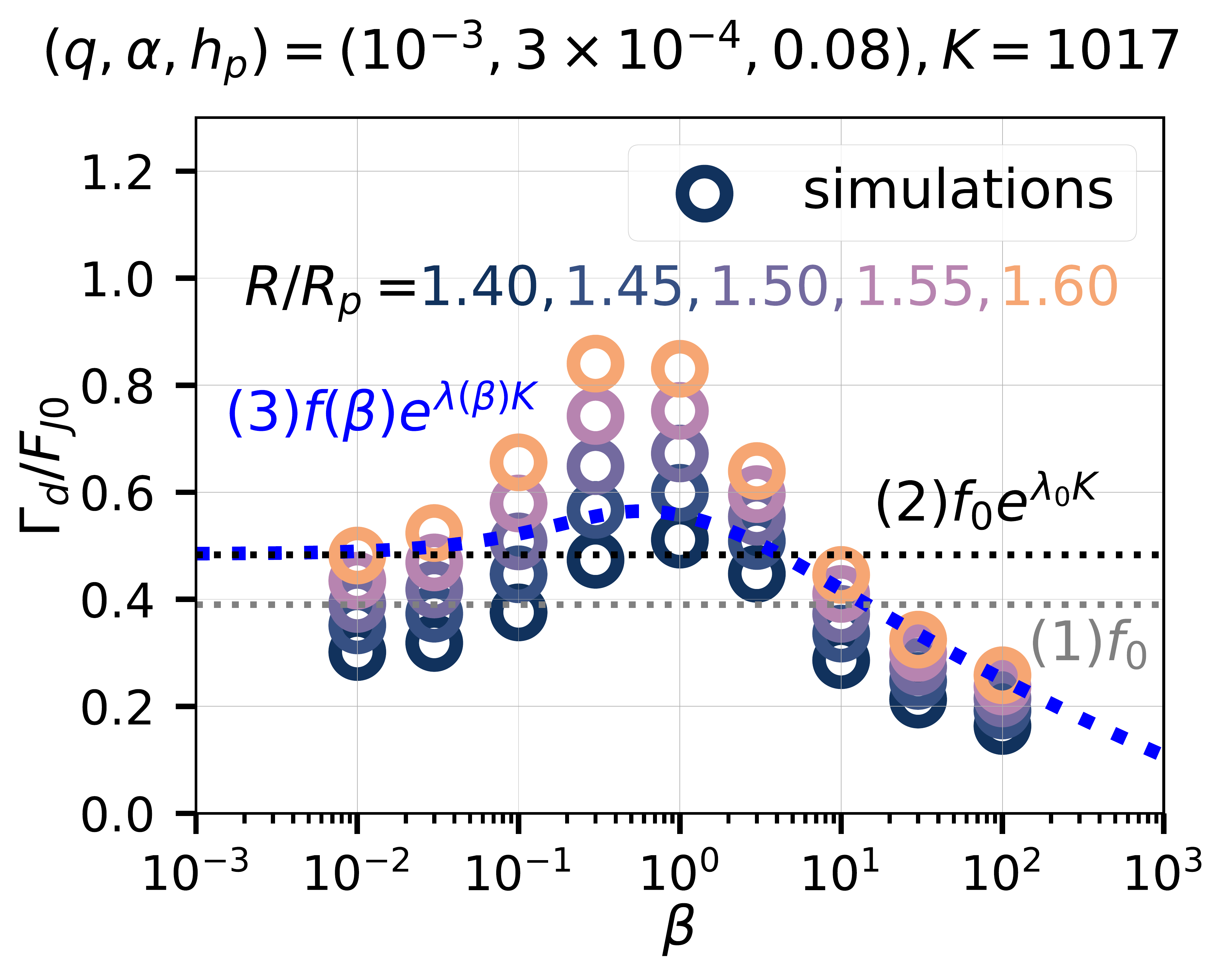}}
\subfigure[]{\includegraphics[width=0.45\textwidth]{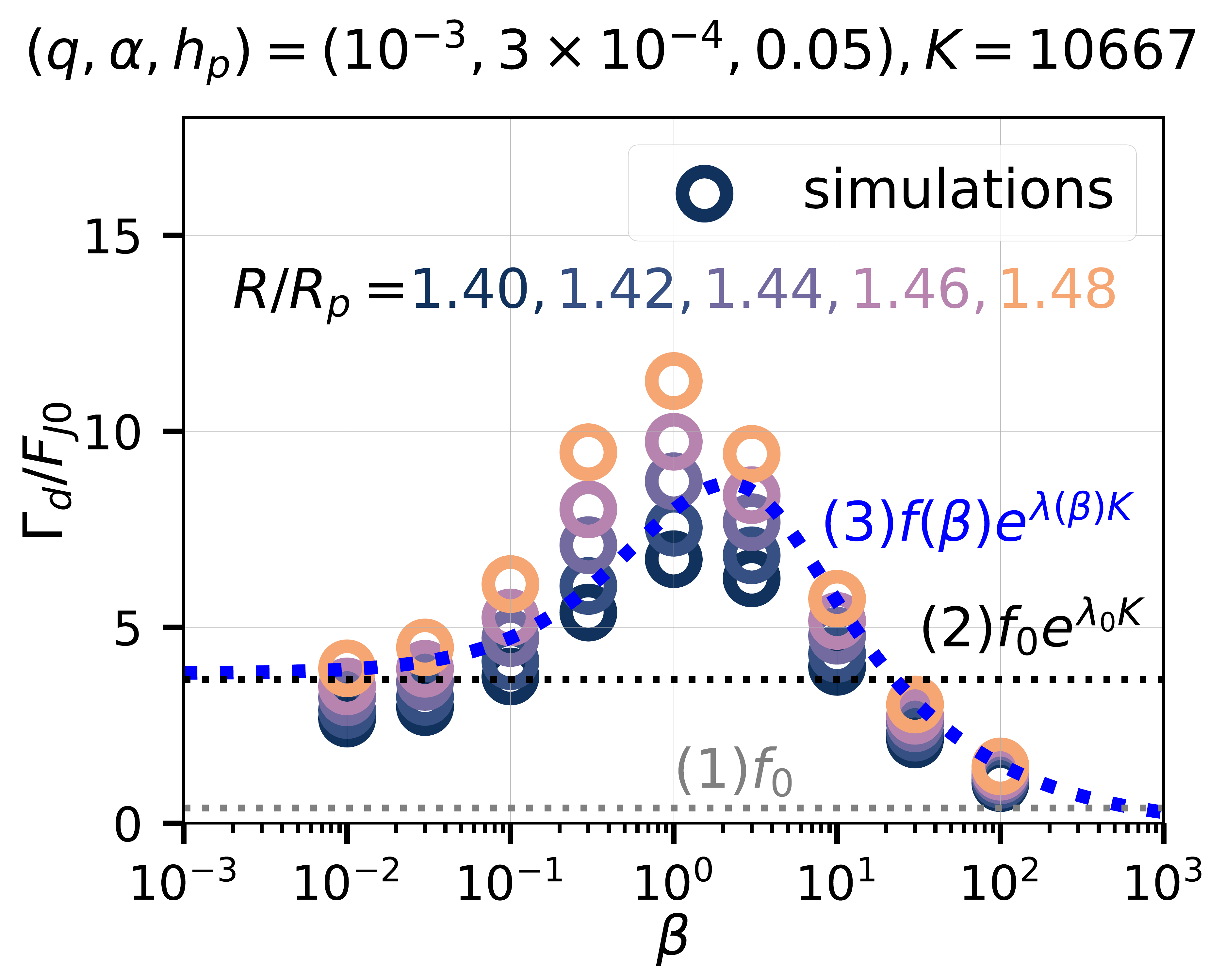}}
\caption{\label{Td_beta}The cumulative deposition torque evaluated near the gap edge with respect to $\beta$.
Circles are $\Gamma_d$ at various $R$ near the gap edge yielded from numerical simulations.
The gray, black, and blue dotted lines symbolizes the deposition factors related to the (1) linear, isothermal (Eq. \ref{K}), (2) nonlinear, isothermal (Eq. \ref{K_nl}), and (3) nonlinear, with cooling (Eq. \ref{K3}) cases, respectively, symbolizing the theoretical deposition torque at the periphery of the gap $(\Gamma_d(R_+))$.
}
\end{figure*}

To verify this physical significance of 
the correction terms in Eq. \ref{K3}, 
we numerically calculate the deposition torque $(\Gamma_d/F_{J0})$ near the gap edge ($R_+$) and compare it with three cases: (1) the linear, isothermal coefficient $f_0$, (2) the coefficient in isothermal disks with nonlinear correction $f_0e^{\lambda_0K}$, and (3) the coefficient with both nonlinear and cooling corrections $f(\beta)e^{\lambda(\beta)K}$. Results are shown in Figure \ref{Td_beta}.  
We observed that the non-linear effect becomes pronounced at $K\gtrsim10^4$, where $f_0e^{\lambda_0K}$ can be $\sim10\times f_0$.
In addition, 
the coefficient $f(\beta)e^{\lambda(\beta)K}$, incorporating both non-linear and cooling adjustments, displays an initial increase followed by a subsequent decrease 
as $\beta$ varies. This behavior mirrors the increasing-decreasing pattern observed in the simulated $\Gamma_d(R_+)$.
Because the gap width varies with $\beta$, the position of $R_+$ will also become closer to the planetary orbit when $\beta\approx1$ and farther when $\beta<1$ or $\beta>1$, which is consistent with Figure \ref{Td_beta}. 

\begin{figure*}[h]
\centering
\subfigure[]{\includegraphics[width=0.45\textwidth]{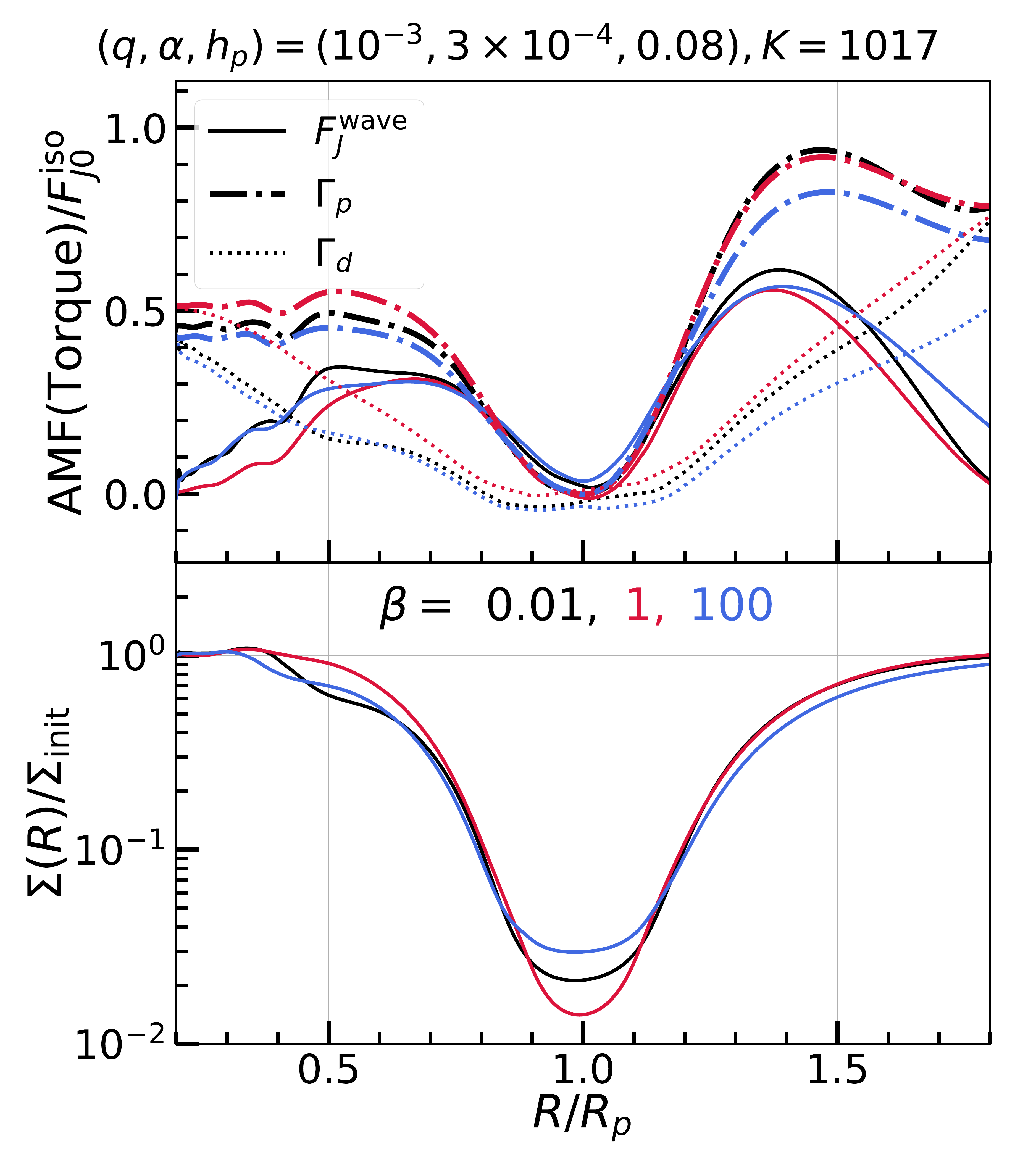}}
\subfigure[]{\includegraphics[width=0.45\textwidth]{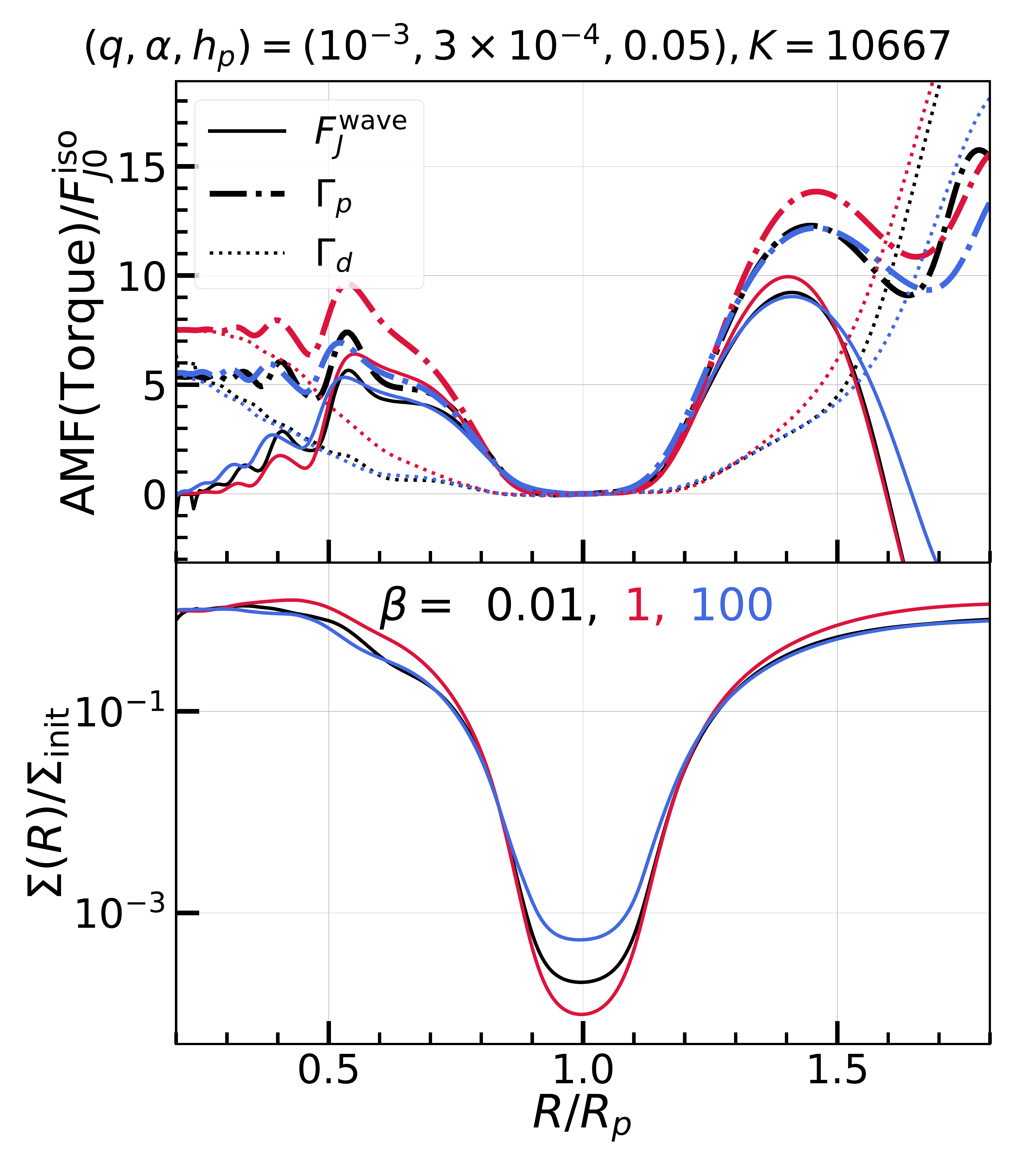}}
\caption{\label{AMF1} The dependence of torques and angular momentum flux on radius and the corresponding radial gap structures under different cooling timescales ($\beta$) in two cases: 
(a) $(q,\alpha,h_p)=(10^{-3},3\times10^{-4},0.08)$, and
(b) $(q,\alpha,h_p)=(10^{-3},3\times10^{-4},0.05)$. 
The top panel shows the radial profiles of AMF and torques. $F_J^{\rm wave}$, $\Gamma_p$, and $\Gamma_d$ are marked in dash-dot, dotted, and solid lines
(see Section \ref{sec:gapdepth} and Eq. (\ref{Tp}), (\ref{Td}), and (\ref{FJwave}) for their definitions),
normalized by $F_{J0}^{\rm iso}=\Sigma_{\rm gap}^{\beta=0.01}q^2h_p^{-3}R_p^4\Omega_p^2$ for the comparability of curves.
The bottom panel shows the azimuthally averaged gap structure. In both cases, $T_d$ with $\beta=1$ (i.e. the red dotted curve) grows the most rapidly, indicating that the disk has its maximum absorption efficiency for wave AMF, and a narrowest and deepest gap will form. }
\end{figure*}

By conducting simulations with different cooling timescales, we observe that the density wave corresponding to $\beta\sim1$ peaks at the position closest to the planet's orbit and fails to propagate further (see Figure \ref{AMF1}). Similar phenomena regarding the dependence of AMF damping on the cooling time scale have been explained in \cite{2020ApJ...904..121M,2020MNRAS.493.2287Z,2022ApJ...930...40S,2023arXiv230514415Z}, among others. However, the excitation torque $\Gamma_p$ in the $\beta=1$ case is slightly larger than in other cases, resulting in a more significant increase in deposition torque $\Gamma_d$ with increasing $R$. As $\Gamma_d(R_+)$ peaks at $\beta=1$, the gap depth $\Sigma_{\text{gap}}$ reaches its minimum value according to Equation \ref{Td2}.

\bibliographystyle{aasjournal}
\bibliography{gap-depth_submitted}{}

\begin{thebibliography}{}
\expandafter\ifx\csname natexlab\endcsname\relax\def\natexlab#1{#1}\fi
\providecommand{\url}[1]{\href{#1}{#1}}
\providecommand{\dodoi}[1]{doi:~\href{http://doi.org/#1}{\nolinkurl{#1}}}
\providecommand{\doeprint}[1]{\href{http://ascl.net/#1}{\nolinkurl{http://ascl.net/#1}}}
\providecommand{\doarXiv}[1]{\href{https://arxiv.org/abs/#1}{\nolinkurl{https://arxiv.org/abs/#1}}}

\bibitem[{{Armitage}(2020)}]{2020apfs.book.....A}
{Armitage}, P.~J. 2020, {Astrophysics of planet formation, Second Edition}

\bibitem[{Bae \& Zhu(2018)}]{Bae_2018}
Bae, J., \& Zhu, Z. 2018, \apj, 859, 119, \dodoi{10.3847/1538-4357/aabf93}

\bibitem[{{Bae} {et~al.}(2017){Bae}, {Zhu}, \&
  {Hartmann}}]{2017ApJ...850..201B}
{Bae}, J., {Zhu}, Z., \& {Hartmann}, L. 2017, \apj, 850, 201,
  \dodoi{10.3847/1538-4357/aa9705}

\bibitem[{{Chen} {et~al.}(2021){Chen}, {Yang}, {Martin}, \&
  {Zhu}}]{2021MNRAS.500.2822C}
{Chen}, C., {Yang}, C.-C., {Martin}, R.~G., \& {Zhu}, Z. 2021, \mnras, 500,
  2822, \dodoi{10.1093/mnras/staa3427}

\bibitem[{{de Val-Borro} {et~al.}(2006){de Val-Borro}, {Edgar}, {Artymowicz},
  {Ciecielag}, {Cresswell}, {D'Angelo}, {Delgado-Donate}, {Dirksen}, {Fromang},
  {Gawryszczak}, {Klahr}, {Kley}, {Lyra}, {Masset}, {Mellema}, {Nelson},
  {Paardekooper}, {Peplinski}, {Pierens}, {Plewa}, {Rice}, {Sch{\"a}fer}, \&
  {Speith}}]{DeValBorro2006}
{de Val-Borro}, M., {Edgar}, R.~G., {Artymowicz}, P., {et~al.} 2006, \mnras,
  370, 529, \dodoi{10.1111/j.1365-2966.2006.10488.x}

\bibitem[{{Dong} \& {Fung}(2017)}]{2017ApJ...835..146D}
{Dong}, R., \& {Fung}, J. 2017, \apj, 835, 146,
  \dodoi{10.3847/1538-4357/835/2/146}

\bibitem[{Dong {et~al.}(2017)Dong, Li, Chiang, \& Li}]{Dong_2017}
Dong, R., Li, S., Chiang, E., \& Li, H. 2017, \apj, 843, 127,
  \dodoi{10.3847/1538-4357/aa72f2}

\bibitem[{Dong {et~al.}(2018)Dong, Li, Chiang, \& Li}]{Dong_2018}
---. 2018, \apj, 866, 110, \dodoi{10.3847/1538-4357/aadadd}

\bibitem[{{Dong} {et~al.}(2017){Dong}, {van der Marel}, {Hashimoto}, {Chiang},
  {Akiyama}, {Liu}, {Muto}, {Knapp}, {Tsukagoshi}, {Brown}, {Bruderer},
  {Koyamatsu}, {Kudo}, {Ohashi}, {Rich}, {Satoshi}, {Takami}, {Wisniewski},
  {Yang}, {Zhu}, \& {Tamura}}]{2017ApJ...836..201D}
{Dong}, R., {van der Marel}, N., {Hashimoto}, J., {et~al.} 2017, \apj, 836,
  201, \dodoi{10.3847/1538-4357/aa5abf}

\bibitem[{{Duffell}(2015)}]{2015ApJ...807L..11D}
{Duffell}, P.~C. 2015, \apjl, 807, L11, \dodoi{10.1088/2041-8205/807/1/L11}

\bibitem[{Duffell(2020)}]{Duffell_2020}
Duffell, P.~C. 2020, \apj, 889, 16, \dodoi{10.3847/1538-4357/ab5b0f}

\bibitem[{Duffell \& Dong(2015)}]{Duffell_2015}
Duffell, P.~C., \& Dong, R. 2015, \apj, 802, 42,
  \dodoi{10.1088/0004-637X/802/1/42}

\bibitem[{Fung \& Chiang(2016)}]{Fung_2016}
Fung, J., \& Chiang, E. 2016, \apj, 832, 105,
  \dodoi{10.3847/0004-637X/832/2/105}

\bibitem[{{Fung} {et~al.}(2014){Fung}, {Shi}, \&
  {Chiang}}]{2014ApJ...782...88F}
{Fung}, J., {Shi}, J.-M., \& {Chiang}, E. 2014, \apj, 782, 88,
  \dodoi{10.1088/0004-637X/782/2/88}

\bibitem[{{Huang} \& {Yu}(2022)}]{2022MNRAS.514.1733H}
{Huang}, S., \& {Yu}, C. 2022, \mnras, 514, 1733,
  \dodoi{10.1093/mnras/stac1464}

\bibitem[{{Isella} {et~al.}(2016){Isella}, {Guidi}, {Testi}, {Liu}, {Li}, {Li},
  {Weaver}, {Boehler}, {Carperter}, {De Gregorio-Monsalvo}, {Manara}, {Natta},
  {P{\'e}rez}, {Ricci}, {Sargent}, {Tazzari}, \&
  {Turner}}]{2016PhRvL.117y1101I}
{Isella}, A., {Guidi}, G., {Testi}, L., {et~al.} 2016, \prl, 117, 251101,
  \dodoi{10.1103/PhysRevLett.117.251101}

\bibitem[{{Kanagawa}(2019)}]{2019ApJ...879L..19K}
{Kanagawa}, K.~D. 2019, \apjl, 879, L19, \dodoi{10.3847/2041-8213/ab2a0f}

\bibitem[{{Kanagawa} {et~al.}(2015{\natexlab{a}}){Kanagawa}, {Muto}, {Tanaka},
  {Tanigawa}, {Takeuchi}, {Tsukagoshi}, \& {Momose}}]{2015ApJ...806L..15K}
{Kanagawa}, K.~D., {Muto}, T., {Tanaka}, H., {et~al.} 2015{\natexlab{a}},
  \apjl, 806, L15, \dodoi{10.1088/2041-8205/806/1/L15}

\bibitem[{{Kanagawa} {et~al.}(2016){Kanagawa}, {Muto}, {Tanaka}, {Tanigawa},
  {Takeuchi}, {Tsukagoshi}, \& {Momose}}]{2016PASJ...68...43K}
---. 2016, \pasj, 68, 43, \dodoi{10.1093/pasj/psw037}

\bibitem[{{Kanagawa} {et~al.}(2017){Kanagawa}, {Tanaka}, {Muto}, \&
  {Tanigawa}}]{2017PASJ...69...97K}
{Kanagawa}, K.~D., {Tanaka}, H., {Muto}, T., \& {Tanigawa}, T. 2017, \pasj, 69,
  97, \dodoi{10.1093/pasj/psx114}

\bibitem[{{Kanagawa} {et~al.}(2015{\natexlab{b}}){Kanagawa}, {Tanaka}, {Muto},
  {Tanigawa}, \& {Takeuchi}}]{2015MNRAS.448..994K}
{Kanagawa}, K.~D., {Tanaka}, H., {Muto}, T., {Tanigawa}, T., \& {Takeuchi}, T.
  2015{\natexlab{b}}, \mnras, 448, 994, \dodoi{10.1093/mnras/stv025}

\bibitem[{{Kanagawa} {et~al.}(2018){Kanagawa}, {Tanaka}, \&
  {Szuszkiewicz}}]{2018ApJ...861..140K}
{Kanagawa}, K.~D., {Tanaka}, H., \& {Szuszkiewicz}, E. 2018, \apj, 861, 140,
  \dodoi{10.3847/1538-4357/aac8d9}

\bibitem[{{Lin} \& {Papaloizou}(1979)}]{1979MNRAS.186..799L}
{Lin}, D.~N.~C., \& {Papaloizou}, J. 1979, \mnras, 186, 799,
  \dodoi{10.1093/mnras/186.4.799}

\bibitem[{{Miranda} \& {Rafikov}(2020{\natexlab{a}})}]{2020ApJ...892...65M}
{Miranda}, R., \& {Rafikov}, R.~R. 2020{\natexlab{a}}, \apj, 892, 65,
  \dodoi{10.3847/1538-4357/ab791a}

\bibitem[{{Miranda} \& {Rafikov}(2020{\natexlab{b}})}]{2020ApJ...904..121M}
---. 2020{\natexlab{b}}, \apj, 904, 121, \dodoi{10.3847/1538-4357/abbee7}

\bibitem[{Muley {et~al.}(2021)Muley, Dong, \& Fung}]{Muley_2021}
Muley, D., Dong, R., \& Fung, J. 2021, \aj, 162, 129,
  \dodoi{10.3847/1538-3881/ac141f}

\bibitem[{Muley {et~al.}(2023)Muley, Fuksman, \&
  Klahr}]{muley2023threetemperature}
Muley, D., Fuksman, J. D.~M., \& Klahr, H. 2023, Three-temperature radiation
  hydrodynamics with PLUTO: Tests and applications to protoplanetary disks.
\newblock \doarXiv{2308.03504}

\bibitem[{{Paardekooper} {et~al.}(2022){Paardekooper}, {Dong}, {Duffell},
  {Fung}, {Masset}, {Ogilvie}, \& {Tanaka}}]{2022arXiv220309595P}
{Paardekooper}, S.-J., {Dong}, R., {Duffell}, P., {et~al.} 2022, arXiv
  e-prints, arXiv:2203.09595, \dodoi{10.48550/arXiv.2203.09595}

\bibitem[{{S{\'a}nchez-Salcedo} {et~al.}(2023){S{\'a}nchez-Salcedo},
  {Chametla}, \& {Chrenko}}]{2023MNRAS.518..439S}
{S{\'a}nchez-Salcedo}, F.~J., {Chametla}, R.~O., \& {Chrenko}, O. 2023, \mnras,
  518, 439, \dodoi{10.1093/mnras/stac2856}

\bibitem[{{Speedie} {et~al.}(2022){Speedie}, {Booth}, \&
  {Dong}}]{2022ApJ...930...40S}
{Speedie}, J., {Booth}, R.~A., \& {Dong}, R. 2022, \apj, 930, 40,
  \dodoi{10.3847/1538-4357/ac5cc0}

\bibitem[{Stone {et~al.}(2020)Stone, Tomida, White, \& Felker}]{Stone_2020}
Stone, J.~M., Tomida, K., White, C.~J., \& Felker, K.~G. 2020, \apjs, 249, 4,
  \dodoi{10.3847/1538-4365/ab929b}

\bibitem[{Teague {et~al.}(2022)Teague, Bae, Benisty, Andrews, Facchini, Huang,
  \& Wilner}]{Teague_2022}
Teague, R., Bae, J., Benisty, M., {et~al.} 2022, \apj, 930, 144,
  \dodoi{10.3847/1538-4357/ac67a3}

\bibitem[{{van der Marel} {et~al.}(2016){van der Marel}, {van Dishoeck},
  {Bruderer}, {Andrews}, {Pontoppidan}, {Herczeg}, {van Kempen}, \&
  {Miotello}}]{2016A&A...585A..58V}
{van der Marel}, N., {van Dishoeck}, E.~F., {Bruderer}, S., {et~al.} 2016,
  \aap, 585, A58, \dodoi{10.1051/0004-6361/201526988}

\bibitem[{{van der Marel} {et~al.}(2015){van der Marel}, {van Dishoeck},
  {Bruderer}, {P{\'e}rez}, \& {Isella}}]{2015A&A...579A.106V}
{van der Marel}, N., {van Dishoeck}, E.~F., {Bruderer}, S., {P{\'e}rez}, L., \&
  {Isella}, A. 2015, \aap, 579, A106, \dodoi{10.1051/0004-6361/201525658}

\bibitem[{{Wang} {et~al.}(2023){Wang}, {Bai}, \& {Lai}}]{2023ApJ...943..175W}
{Wang}, H.-Y., {Bai}, X.-N., \& {Lai}, D. 2023, \apj, 943, 175,
  \dodoi{10.3847/1538-4357/acac77}

\bibitem[{{Yen} {et~al.}(2019){Yen}, {Gu}, {Hirano}, {Koch}, {Lee}, {Liu}, \&
  {Takakuwa}}]{2019ApJ...880...69Y}
{Yen}, H.-W., {Gu}, P.-G., {Hirano}, N., {et~al.} 2019, \apj, 880, 69,
  \dodoi{10.3847/1538-4357/ab29f8}

\bibitem[{Yen {et~al.}(2016)Yen, Liu, Gu, Hirano, Lee, Puspitaningrum, \&
  Takakuwa}]{Yen_2016}
Yen, H.-W., Liu, H.~B., Gu, P.-G., {et~al.} 2016, \apjl, 820, L25,
  \dodoi{10.3847/2041-8205/820/2/L25}

\bibitem[{Zhang {et~al.}(2021)Zhang, Booth, Law, Bosman, Schwarz, Bergin,
  Öberg, Andrews, Guzmán, Walsh, Qi, van~’t Hoff, Long, Wilner, Huang,
  Czekala, Ilee, Cataldi, Bergner, Aikawa, Teague, Bae, Loomis, Calahan,
  Alarcón, Ménard, Gal, Sierra, Yamato, Nomura, Tsukagoshi, Pérez, Trapman,
  Liu, \& Furuya}]{Zhang_2021}
Zhang, K., Booth, A.~S., Law, C.~J., {et~al.} 2021, \apjs, 257, 5,
  \dodoi{10.3847/1538-4365/ac1580}

\bibitem[{{Zhang} \& {Zhu}(2020)}]{2020MNRAS.493.2287Z}
{Zhang}, S., \& {Zhu}, Z. 2020, \mnras, 493, 2287,
  \dodoi{10.1093/mnras/staa404}

\bibitem[{Zhang {et~al.}(2018)Zhang, Zhu, Huang, Guzmán, Andrews, Birnstiel,
  Dullemond, Carpenter, Isella, Pérez, Benisty, Wilner, Baruteau, Bai, \&
  Ricci}]{Zhang_2018}
Zhang, S., Zhu, Z., Huang, J., {et~al.} 2018, \apjl, 869, L47,
  \dodoi{10.3847/2041-8213/aaf744}

\bibitem[{Zhu {et~al.}(2015)Zhu, Dong, Stone, \& Rafikov}]{Zhu_2015}
Zhu, Z., Dong, R., Stone, J.~M., \& Rafikov, R.~R. 2015, \apj, 813, 88,
  \dodoi{10.1088/0004-637X/813/2/88}

\bibitem[{{Ziampras} {et~al.}(2020){Ziampras}, {Ataiee}, {Kley}, {Dullemond},
  \& {Baruteau}}]{2020A&A...633A..29Z}
{Ziampras}, A., {Ataiee}, S., {Kley}, W., {Dullemond}, C.~P., \& {Baruteau}, C.
  2020, \aap, 633, A29, \dodoi{10.1051/0004-6361/201936495}

\bibitem[{{Ziampras} {et~al.}(2023){Ziampras}, {Nelson}, \&
  {Rafikov}}]{2023arXiv230514415Z}
{Ziampras}, A., {Nelson}, R.~P., \& {Rafikov}, R.~R. 2023, arXiv e-prints,
  arXiv:2305.14415, \dodoi{10.48550/arXiv.2305.14415}

\end{thebibliography}
\end{CJK*}
\end{document}